\documentclass[%
reprint,
superscriptaddress,
amsmath,amssymb,
aps,
pra,
10pt]{revtex4-2}

\usepackage{graphicx, subfigure}
\usepackage{dcolumn}
\usepackage{bm}
\usepackage[colorlinks,citecolor=blue]{hyperref}
\usepackage{physics}
\usepackage{dsfont}
\usepackage{amsmath}
\usepackage{braket}
\usepackage{amsfonts}
\usepackage{amssymb}
\usepackage{siunitx}
\usepackage{easyReview}
\usepackage{multirow}
\usepackage{footnote}

\begin{document}
	
   	\title{Strong Noninertial Radiative Shifts in Atomic Spectra at Low Accelerations}
	\author{Navdeep Arya}
	\email{navdeeparya.me@gmail.com}
    \email{navdeep.arya@fysik.su.se}
	\affiliation{Department of Physical Sciences, Indian Institute of Science Education \& Research (IISER) Mohali, Sector 81 SAS Nagar, Manauli PO 140306 Punjab India}
    \affiliation{Department of Physics, Stockholm University, Roslagstullsbacken 21, 106 91 Stockholm, Sweden}
	\author{D. Jaffino Stargen}
 \email{jaffinostargend@gmail.com}
    \email{jaffino@joyuniversity.edu.in}
	\affiliation{Quantum  \& Precision Measurements Group, Department of Mechanical Engineering, Massachusetts Institute of Technology, 77 Massachusetts Ave, Cambridge, MA 02139, USA}
    \affiliation{Department of Physical Sciences, School of Arts \& Natural Sciences, Joy University,
    Vadakangulam (Near Kanyakumari), Tamilnadu -- 627116, India}
	\author{Kinjalk Lochan}
    \email{kinjalk@iisermohali.ac.in}
	\affiliation{Department of Physical Sciences, Indian Institute of Science Education \& Research (IISER) Mohali, Sector 81 SAS Nagar, Manauli PO 140306 Punjab India}
	\author{Sandeep K.~Goyal}
	\email{skgoyal@iisermohali.ac.in}
	\affiliation{Department of Physical Sciences, Indian Institute of Science Education \& Research (IISER) Mohali, Sector 81 SAS Nagar, Manauli PO 140306 Punjab India}
	
	\date{\today}
	
	\begin{abstract}
		Despite numerous proposals investigating various properties of accelerated detectors in different settings, detecting the Unruh effect remains challenging due to the typically weak signal at achievable accelerations.
		For an atom with frequency gap $\omega_0$, accelerated in free space, significant acceleration-induced modification of properties like transition rates and radiative energy shifts requires accelerations of the order of $\omega_0 c$.
		In this paper, we make the case for a suitably modified density of field states to be complemented by a judicious selection of the system property to be monitored.
		We study the radiative energy-level shift in inertial and uniformly accelerated atoms coupled to a massless quantum scalar field inside a cylindrical cavity.
		Uniformly accelerated atoms experience thermal correlations in the inertial vacuum, and the radiative shifts are expected to respond accordingly.
		We show that the noninertial contribution to the energy shift can be isolated and significantly enhanced relative to the inertial contribution by suitably modifying the density of field modes inside a cylindrical cavity. Moreover, we demonstrate that monitoring the radiative energy shift, as compared to transition rates, allows us to reap a stronger purely-noninertial signal. We find that a purely-noninertial radiative shift as large as 50 times the inertial energy shift can be obtained at small, experimentally achievable accelerations ($ a \sim 10^{-9} \omega_{0} c$) if the cavity's radius $R$ is specified with a relative precision of $\delta R/R_{0} \sim 10^{-7}$. Given that radiative shifts for inertial atoms have already been measured with high accuracy, we argue that the radiative energy-level shift is a promising observable for detecting Unruh thermality with current technology. 
	\end{abstract}
	
	\maketitle
	
	\section{Introduction}
The Unruh effect is a key predication at the interface of the theory of relativity and quantum theory. It states that a uniformly accelerated observer perceives the inertial vacuum as a thermal state at a temperature proportional to the observer's acceleration~\cite{Fulling1973,Davies1975,Unruh1976}. A large number of proposals~\cite{Rogers1988,Tajima1999,Vanzella2001,scully2003,fuentes2010,Martinez2013,Barshay1978,*Barshay1980,*Kharzeev2006} for the detection of the Unruh effect still await fruition as the thermal signature at the achievable accelerations is very feeble under traditional settings. In order to have any verification worthy effects, even with the most sophisticated experimental setups in the foreseeable future, accelerations on the order of $10^{20}~{\rm m/s^2}$ are required.

There have been proposals exploiting the high sensitivity of probes such as entanglement and geometric phase~\cite{Salton:2014jaa,Eduardo2011,HuYu2012} to relax extreme requirements and obtain potential noninertial signatures under moderate conditions.
Formidable as these proposals may be, they still demand accelerations and expertise much beyond what can be made achievable in near future. However, these schemes do put forward an idea that noninertial motion of a system might affect different system properties differently~\cite{Arya2022,Arya2023}. Additionally, different system properties might differ with respect to the level of maturity of the experimental techniques required for their measurement.

Furthermore, modifying the field’s boundary conditions to better resolve noninertial contributions at low accelerations has been identified as a promising approach~\cite{Lochan:2019,Vriend2021}. For instance, inside a cylindrical cavity, it has been argued that the noninertial contribution to the emission rate of an atom uniformly accelerated along the cavity's axis can be made significantly dominant over the inertial contribution even at low accelerations provided that the cavity's geometry is specified precisely enough~\cite{Jaffino2022}.

 Combining these two ideas, we look for a system property in a cavity environment which not only captures noninertial effects more efficiently but also has precedents of observations at least when dealing with inertial atoms which can be adapted to noninertial setups. In this spirit, we study the radiative energy shift in an inertial and a uniformly accelerated two-level atom (hereafter referred to as a \textit{Rindler atom}) coupled to a real massless quantum scalar field inside a long cylindrical cavity. The radiative energy shift in atomic energy levels arises due to the atomic electron's coupling to an external quantum field and has been measured with great precision in inertial settings using different experimental methods ~\cite{Hagley1994,Weitz1994,Berkeland1995,Bezginov2019}. The radiative shifts in the atomic spectra hold interest due to intense experimental activity surrounding atomic spectroscopy and the resultant high-precision measurements of the spectral lines~\cite{CODATA2018,Beyer2016,Beyer2017,Fleurbaey2018,Bezginov2019,Grinin2020,Skinner2024}. Additionally, the radiative energy shifts in inertial atoms even under modified density of field modes (e.g., by introducing a mirror~\cite{Wilson2003} or parallel metal plates~\cite{Marrocco1998} or by placing the atom inside a confocal resonator~\cite{Heinzen1987}) have also been measured with precision. 

The radiative energy shifts are essentially controlled by the field correlations perceived by the atom which, in turn, depend on the state of motion of the atom and the density of field modes. For instance a uniformly accelerating observer perceives the vacuum fluctuations akin to thermal correlations. Therefore, the radiative energy shifts should be different for an inertial and a noninertially moving atom. 
Our focus in this work is to study the behavior of the radiative energy shift in an inertial and a Rindler atom  inside a cavity where the density of modes can be controlled by modifying the cavity geometry.
Note, for instance, that in Hydrogen atom the optical and microwave transitions have been measured with precision in the ranges $10^{-12} - 10^{-11}$ and $10^{-6} - 10^{-5}$, respectively~\cite{CODATA2018,Karr2020}.
Availability of such precisely measured transition lines promises to be instrumental in determining various corrections to the spectral lines, including the ones attributable to the atom's acceleration~\cite{Arya2023}.

In this work, we consider a cylindrical cavity in which the density of field modes is modified such that the emission rate of an inertial atom rises abruptly when the cavity is tuned such that the frequency of a transverse cavity mode matches with the atom's resonance frequency (hereafter called the \textit{resonance points} of the atom-cavity system) and falls off quickly away from a resonance point~\cite{Jaffino2022}. The emission rate of a Rindler atom, on the other hand, is finite at the resonance points and falls off slowly (as compared to the inertial emission rate) away from the resonance points with an oscillatory behavior. By tuning the cavity slightly away from a resonance point, the noninertial emission rate of the Rindler atom can be made dominant over the atom's inertial emission rate. Achieving the same dominance at lower accelerations requires specification of the cavity radius with a correspondingly higher precision. Thus, the requirement of larger accelerations can be traded for a greater precision in the cavity design~\cite{Jaffino2022}. Since the radiative energy shifts of the atomic levels are caused by the emission and absorption of virtual photons by an atomic electron~\cite{API_Tannoudji,loudon2000quantum}, modified density of field states inside a cylindrical cavity will lead to modifications in the energy shifts as well~\cite{Barton1970,*Barton1979,*Barton1987,Billaud2013}. In fact, as we shall show, at low accelerations a purely-noninertial energy shift\textemdash orders of magnitude larger than the inertial energy shift\textemdash can be obtained inside an appropriately designed cavity.

In this paper, we show that smaller accelerations lead to a larger energy level shift but require specification of the cavity's radius with a greater precision. We demonstrate that as compared to the atom's transition rates, monitoring radiative energy shift in a Rindler atom's spectra allows one to reap a stronger purely-noninertial signal. For an atom with proper (i.e., measured in its rest frame) frequency-gap $\omega_{0}$, we show that a purely-noninertial radiative shift, as large as 50 times the inertial energy shift can be obtained at small, experimentally achievable accelerations ($ a \sim 10^{-9} \omega_0 c$) if the cavity's radius $R$ is specified with a relative precision of $\delta R/R_{0} \sim 10^{-7}$. We point out relevant experiments dealing with inertial atoms which can be adapted to the noninertial setup under consideration here.

This paper is organized as follows. In Sec.~\ref{Sec:Quantized scalar field inside a cylindrical cavity}, we start by discussing quantization of a scalar field inside a long cylindrical cavity from the perspective of an inertial observer. This discussion is followed in Sec.~\ref{Sec: RES-background} by a summary of the Dalibard, Dupont-Roc, and Cohen-Tannoudji (DDC) formalism for the determination of radiative energy shifts in a small system coupled weakly to a large reservoir. Thereafter, Secs.~\ref{Sec: RESIn-results} and \ref{Sec: RESnin-results} present computations of the radiative energy shift in inertial and Rindler atoms, respectively, and discuss the results. We conclude with Sec.~\ref{Sec: conclusion}.

	\section{Background}\label{background}
	\subsection{Quantized scalar field inside a cylindrical cavity}\label{Sec:Quantized scalar field inside a cylindrical cavity}
\begin{figure}
	\centering
	\includegraphics[width=0.9\linewidth]{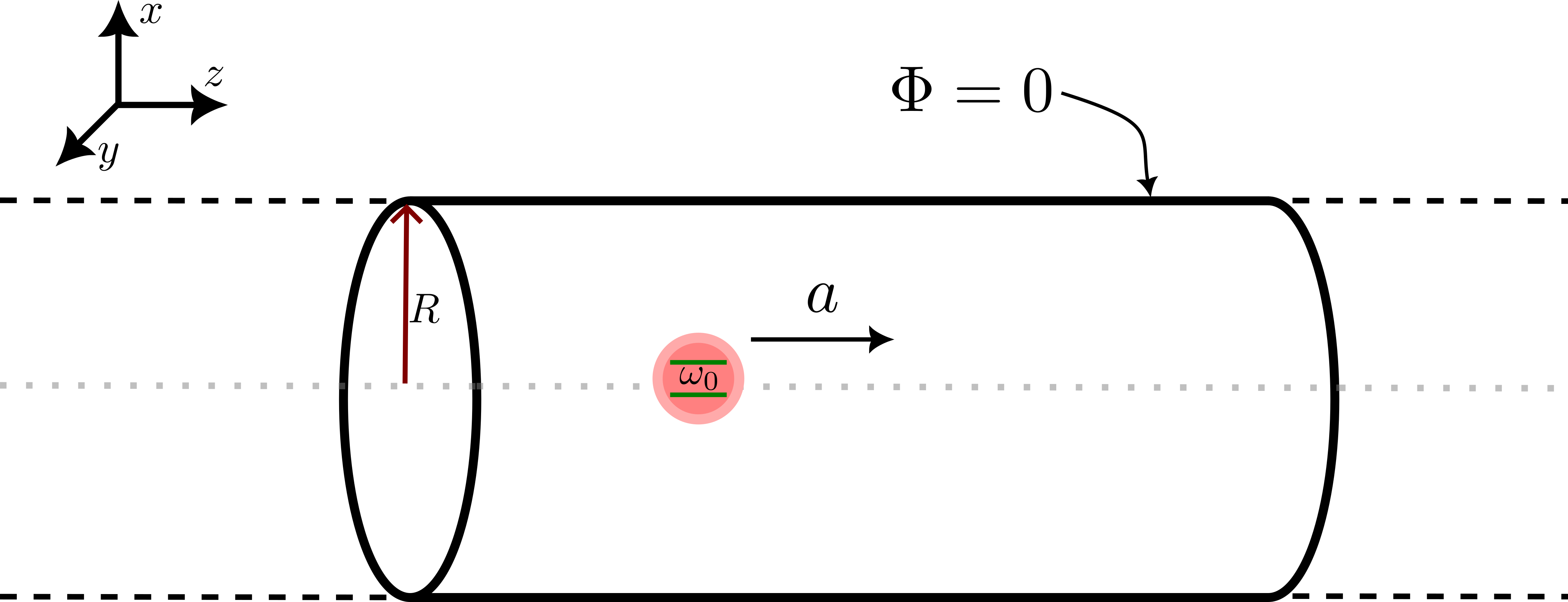}
	\caption{A two-level atom, with frequency gap $\omega_0$, moving along the axis of a long cylindrical cavity of length $L \gg c/\omega_0$ and radius $R$. For an inertial atom $a = 0$, but for a Rindler atom $a \neq 0$ but a constant. The field $\Phi(t,\vb{x})$ is assumed to satisfy vanishing Dirichlet boundary conditions at the cavity walls.}
	\label{fig:cycav}
\end{figure}

	In this work, we study a two-level atom, having proper frequency gap $\omega_0$, interacting with a quantized massless real scalar field $\Phi$ inside a long cylindrical cavity of length $L \gg c/\omega_0$ and radius $R$. Therefore, we start by discussing the quantization of the scalar field inside such a cavity, with the field satisfying vanishing Dirichlet boundary conditions at the cavity walls~[see Fig.~\ref{fig:cycav}]. Assume the cavity's axis to be aligned with the $z$-direction of the coordinate system.
	The inertial spacetime line element in cylindrical coordinates $(t,\rho,\theta, z)$ can be expressed as
	\begin{equation}\label{metric}
		\dd{s}^2 = - \dd{t}^2 + \dd{\rho}^2 + \rho^2 \dd{\theta}^2 + \dd{z}^2 \equiv g_{\mu \nu}(\tilde{x}) \dd{x}^{\mu} \dd{x}^{\nu}
	\end{equation}
    where $\mu, \nu = 0,1,2,3$; $g_{\mu \nu}(\tilde{x})$ are components of the spacetime metric; and $\tilde{x}$ is a spacetime event. Solving the Klein-Gordon equation
    \begin{equation}
    	\frac{1}{\sqrt{-\tilde{g}}} \partial^{\mu} \left(\sqrt{-\tilde{g}}~ g_{\mu \nu} \partial^{\nu} \phi\right) = 0,
    \end{equation}
    where $\tilde{g} \equiv \det(g_{\mu \nu})$, in the spacetime described by Eq.~\eqref{metric} and imposing the vanishing
    Dirichlet boundary conditions on the field modes at the cavity walls, we obtain a complete set $\{f_{k_z;mn}\}$ of orthonormal solutions with
    \begin{equation}
    	f_{k_z;mn}(\tilde{x}) = \frac{1}{2\pi R \sqrt{\omega_k}} \frac{J_m(\xi_{mn}\rho/R)}{J_{\abs{m}+1}(\xi_{mn})} e^{-i\omega_k t} e^{im\theta} e^{i k_z z},
    \end{equation}
   where $\omega_k^2 \equiv k^2_z + (\xi_{mn}/R)^2$, $k_z \in (-\infty, \infty)$, $\xi_{mn}$ is the $n$-th zero of the Bessel function $J_m(x)$, i.e., $J_m(\xi_{mn})=0$; and $m,n$ are integers such that $-\infty < m < \infty, ~ 1 \leq n < \infty$. Following the usual canonical quantization procedure, the quantized scalar field inside the cylindrical cavity can be expressed as:
   \begin{multline}\label{field-inertial}
   	\Phi(\tilde{x}) = \sum_{m=-\infty}^{\infty} \sum_{n=1}^{\infty} \int_{-\infty}^{\infty} \dd{k_z} \Big(a_{k_z;mn} f_{k_z;mn}(\tilde{x}) \\
   	+ a^{\dagger}_{k_z;mn} f^*_{k_z;mn}(\tilde{x})\Big),
   \end{multline}
with the field annihilation and creation operators satisfying the following commutation relations
\begin{subequations}
	\begin{align}
		[a_{k_z;mn}, a^{\dagger}_{k_z';m'n'}] &= \delta_{mm'} \delta_{nn'} \delta(k_z - k'_z), \\
		[a_{k_z;mn}, a^{}_{k_z';m'n'}] &= 0.
	\end{align}
\end{subequations}
The inertial vacuum state $\ket{0}$ of the cavity field is defined as $a_{k_z;mn} \ket{0} = 0$, for all field modes $ \{k_z;mn\}$. The vacuum state two-point correlator $\mathcal{W}(\tilde{x}(\tau),\tilde{x}(\tau')) \equiv \expval{\Phi(\tilde{x}(\tau))\Phi(\tilde{x}(\tau'))}{0}$ of the field can be computed in a straightforward manner to be:
\begin{multline}\label{Wightman}
\mathcal{W}(x,x') = \frac{1}{2(\pi R)^2} \sum_{m=-\infty}^{\infty} \sum_{n=1}^{\infty} \frac{J_m(\xi_{mn}\rho/R) J_m(\xi_{mn}\rho'/R)}{J^2_{\abs{m}+1}(\xi_{mn})}\\
	\times \int_{0}^{\infty} \frac{\dd{\omega_k}}{\sqrt{\omega^2_k - (\xi_{mn}/R)^2}} e^{-i\omega_k(t-t')} e^{im(\theta - \theta')} e^{i k_z (z-z')}.
\end{multline}
Note that $\mathcal{W}(\tilde{x},\tilde{x}')$ receives a large contribution from field modes with frequency $\omega_k \rightarrow \xi_{mn}/R$. We will call the field modes with $\xi_{mn}/R = \omega_0$ as the \textit{atom-cavity resonance points}. When the cavity radius is such that $R \omega_0 = \xi_{mn} + \epsilon$, where $\epsilon$ is some real number, the cavity will be said to be \textit{detuned} by an amount $\epsilon$. 

\subsection{Radiative energy level shift}\label{Sec: RES-background}
As already stated, our goal in this work is to study radiative energy shift in inertial as well as Rindler atoms inside a cylindrical cavity. Accordingly, in this section we provide a summary of the Dalibard, Dupont-Roc, and Cohen-Tannoudji (DDC) formalism~\cite{DDC1984} for determination of radiative energy shifts in a small system ${\rm S}$ which is weakly coupled to a large reservoir ${\rm R}$. In our case, the small system ${\rm S}$ is a two-level atom and the reservoir ${\rm R}$ is a massless real scalar quantum field.
	
The total Hamiltonian of the composite system ${\rm S+R}$ is given as
\begin{equation}
		H = H_S + H_R + H_I,
	\end{equation}
where $H_S$ and $H_R$ are the system and reservoir Hamiltonians, respectively, and $H_I$ is the system-reservoir interaction Hamiltonian. We consider an interaction Hamiltonian of the form~\cite{DDC1984, breuer2002}
\begin{equation}
	H_I = - g \sum_{i} S_i \otimes R_i,
\end{equation}
where $g$ is a small coupling constant and $S_i$ and $R_i$ are hermitian operators of the system and reservoir, respectively. The time evolution of a system or reservoir operator $O(\tau)$ is governed by the Heisenberg equation
\begin{equation}\label{HeisenbergEq}
	\begin{split}
		\derivative{O(\tau)}{\tau} &= i \comm{H(\tau)}{O(\tau)},
	\end{split}
\end{equation}
which can be solved as a power series in $g$. The solution can formally be written as~\cite{DDC1984}
\begin{equation*}
	O(\tau) = O^{\mathrm{f}}(\tau) + O^{\mathrm{s}}(\tau),
\end{equation*} 
where $O^{\mathrm{f}}(\tau)$ and $ O^{\mathrm{s}}(\tau)$, respectively, are the solutions of Eq.~\eqref{HeisenbergEq} to the zeroth order and to first and higher orders in $g$. Here superscripts `${\rm f}$' and `${\rm s}$' denote `free' and `source' contributions, respectively. We assume the system and the reservoir to be in an uncorrelated state $\rho(\tau)$ at the initial time $\tau_0$, that is, $\rho(\tau_0) = \rho_S(\tau_0) \otimes \rho_R(\tau_0)$. The reservoir-averaged rates of variation of a system observable can be obtained as a perturbative series in $g$. In the reservoir-averaged rates of variation so obtained, one can identify a part whose physical effect is that of an effective Hamiltonian which leads to the modification of the system's energy spectrum, and a non-Hamiltonian part which controls the dissipative dynamics of the system observables. The effective Hamiltonian gets contributions from the reservoir fluctuations (rf) and self-reaction (sr) so that the Hamiltonian part of the system's evolution is now described by $H_S(\tau) + (H_{\rm eff})^{\rm rf} + + (H_{\rm eff})^{\rm sr}$. The resulting energy shift in an unperturbed energy level of the system is given by expectation value of the effective Hamiltonian in that energy eigenstate. Here, we are interested in the role of acceleration in the modification of the energy spectrum of a Rindler atom coupled to a quantum scalar field inside a long cylindrical cavity. Hereafter, `system' and `reservoir' may be taken to mean an `atom' and a `field-cavity setup', respectively.

Let $\{\ket{a}\}$ be the eigenstates of $H_{\rm S}$ with eigenvalues $\{\varepsilon_a\}$. If $\rho_R(\tau_0)$ is a stationary state of the reservoir, that is, if $\comm{H_R}{\rho_R(\tau_0)}=0$, then the reservoir two-point correlation functions depend only on the time difference $u \equiv \tau- \tau'$ corresponding to the two points~\cite{DDC1984,breuer2002}. Further, we restrict ourselves to time scales much larger than the reservoir correlation time scale  $\tau_c$~\cite{DDC1984}.
Then, the energy shifts $\delta E^{\mathrm{(rf)}}_a$ and $\delta E^{\mathrm{(sr)}}_a$ produced respectively by the reservoir fluctuations and self reaction are given, to second order in $g$, by~\cite{DDC1984}:
\begin{subequations}\label{shifts}
	\begin{equation}\label{shiftRF}
		(\delta E_a)_{\mathrm{rf}} = - \frac{g^2}{2} \sum_{i,j} \int_{0}^{\infty} \dd{u} C^{\mathrm{(R)}}_{ij}(u) \chi_{ij}^{\mathrm{(S,a)}}(u),
	\end{equation}
and
\begin{equation}\label{shiftSR}
	(\delta E_a)_{\mathrm{sr}} = - \frac{g^2}{2} \sum_{i,j} \int_{0}^{\infty} \dd{u} \chi^{\mathrm{(R)}}_{ij}(u) C_{ij}^{\mathrm{(S,a)}}(u).
\end{equation}
\end{subequations}
Here
\begin{subequations}\label{statfxns}
	\begin{equation}
		C^{\mathrm{(R)}}_{ij}(u) \equiv \frac{1}{2} \mathrm{Tr_R}\left(\rho_R(\tau_0) \left\{R^{\rm f}_i(\tau_0), R^{\rm f}_j(\tau_0 - u)\right\}\right),
	\end{equation}
\begin{equation}
	\chi^{\mathrm{(R)}}_{ij}(u) \equiv i \mathrm{Tr_R}\left(\rho_R(\tau_0) \left[R^{\rm f}_i(\tau_0), R^{\rm f}_j(\tau_0 - u)\right]\right),
\end{equation}
\begin{equation}
 C^{\mathrm{(S,a)}}_{ij}(u) \equiv \frac{1}{2} \bra{a} \left\{S^{\rm f}_i(\tau_0), S^{\rm f}_j(\tau_0 - u)\right\}\ket{a},
\end{equation}
and
\begin{equation}
	\chi^{\mathrm{(S,a)}}_{ij}(u) \equiv i \bra{a} \left[S^{\rm f}_i(\tau_0), S^{\rm f}_j(\tau_0 - u)\right]\ket{a},
\end{equation}
\end{subequations}
are known as the statistical functions of the reservoir (superscript ${\rm R}$) and the system (superscript ${\rm S}$), respectively. 
\subsubsection*{Two-level atom}
Next, we specialize to a two-level detector. We denote the excited and the ground states of the detector by $\ket{e}$ and $\ket{g}$, respectively. The free Hamiltonian of the detector is $H_{\rm S} = \omega_0 \sigma_{z}/2$, where $\sigma_z$ is one of the Pauli matrices. The detector has a monopole moment $\mu(\tau)$ which couples linearly to the scalar field. That is, the detector-field interaction Hamiltonian is given by~\cite{Unruh1976,einstein1979general,Martinez2020,Martinez2021}
\begin{equation}
	H_I(\tau) = - g \mu(\tau) \Phi(\tilde{x}(\tau)),
\end{equation} 
where $\tau$ is the proper time of the detector and $\tilde{x}(\tau)$ denotes the detector's trajectory in spacetime. This Hamiltonian is a good model for light-matter interaction when no exchange of angular momentum is involved~\cite{Martinez2013}. In the interaction picture, the detector's monopole moment operator is given by
\begin{equation}
	\mu(\tau) = \sigma_- e^{-i \omega_0 \tau} + \sigma_+ e^{i \omega_0 \tau},
\end{equation}
where $\sigma_- = \dyad{g}{e}$ and $\sigma_+ = \dyad{e}{g}$.
We assume that the cavity field is in the inertial vacuum state $\ket{0}$. Therefore, the field and the detector statistical functions take the form~[refer to Eqs.~\eqref{statfxns}]
\begin{subequations}\label{StatFxns2}
	\begin{equation}\label{FieldCorrFxn}
		C^{\mathrm{(R)}}(u) = \frac{1}{2} \bra{0} \left\{\Phi^{\rm f}(\tau_0), \Phi^{\rm f}(\tau_0 - u)\right\} \ket{0},
	\end{equation}
	\begin{equation}\label{FieldSuscep}
		\chi^{\mathrm{(R)}}(u) = i \bra{0} \left[\Phi^{\rm f}(\tau_0), \Phi^{\rm f}(\tau_0 - u)\right]\ket{0},
	\end{equation}
\begin{equation}
		C^{\mathrm{(S,b)}}(u) = \frac{1}{2} \sum_{d} \abs{\bra{b}\mu(0)\ket{d}}^2 \left(e^{i\omega_{bd} u} + e^{- i\omega_{bd} u}\right).
	\end{equation}
	and
 \begin{equation}
	\chi^{\mathrm{(S,b)}}(u) = i \sum_{d} \abs{\bra{b}\mu(0)\ket{d}}^2 \left(e^{i\omega_{bd} u} - e^{- i\omega_{bd} u}\right).
\end{equation}
\end{subequations}
In the above expressions, $\omega_{bd} \equiv \omega_b - \omega_d$ is the frequency difference between atomic levels $\ket{b}$ and $\ket{d}$. The energy level shift in the two-level detector can be obtained by substituting Eqs.~\eqref{StatFxns2} in Eqs.~\eqref{shifts}.
\begin{figure}
	\centering
	\includegraphics[width=0.95\linewidth]{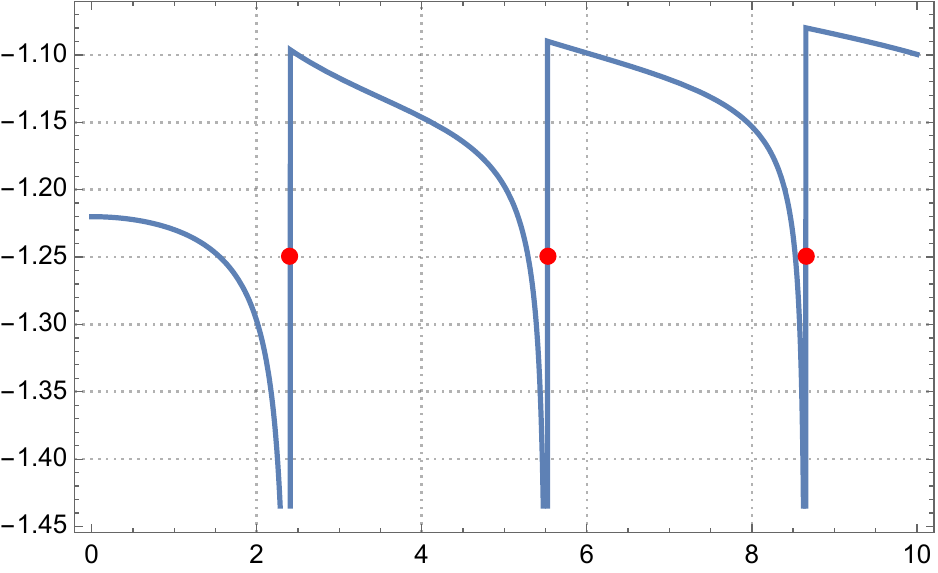}
	\caption{The variation of the energy level shift in an inertial atom moving along the axis of a cylindrical cavity as a function of the cavity-detuning parameter $R\omega_0$. For reference, red dots have been placed at $R\omega_0 = \xi_{0n}, n= 1,2,3$, that is, at the first three atom-cavity resonance points.
    Refer to Eq.~\eqref{LSInertial} and the discussion that follows.}
	\label{fig:LSrest}
\end{figure}
\section{Radiative shift of an atom's spectrum inside a cylindrical cavity}\label{Sec: RES-results}
Now, we consider an inertial and a Rindler atom inside the cylindrical cavity moving along the cavity's axis. We assume the cavity to be a perfect cavity, that is, the field modes satisfy vanishing Dirichlet boundary conditions at the cavity walls~[see Sec.~\ref{Sec:Quantized scalar field inside a cylindrical cavity}].
\subsection{Radiative energy level shift in an inertial atom}\label{Sec: RESIn-results}
Consider an atom moving inertially on the trajectory ${\tilde x}^{\nu}(\tau) = \left(\gamma \tau, 0, 0, \gamma v_0 \tau\right)$ along the axis of a long cylindrical cavity of radius $R$. Here, $\gamma = (1 - v_0^2)^{-1/2}$ is the Lorentz factor. The vacuum field correlation function can be obtained, using Eq.~\eqref{Wightman}, to be
\begin{equation}\label{InertialWight1}
	\begin{split}
		&\bra{0} \Phi(\tilde{x}(\tau_2)) \Phi(\tilde{x}(\tau_1))\ket{0} = \frac{1}{(2 \pi R)^2} \\
		& \times \sum_{m=-\infty}^{\infty} \sum_{n=1}^{\infty} \frac{J^2_m(\xi_{mn} \rho_0/R)}{J^2_{\abs{m}+1}(\xi_{mn})} \int_{-\infty}^{\infty} \frac{\dd{k_z}}{\omega_k} e^{-i \gamma (\omega_k - v_0 k_z) (\tau_2-\tau_1)}.
	\end{split}
\end{equation}
Using the expression for the boosted field momentum $k'^{\mu} = \left(\partial x'^{\mu}/\partial x^{\nu}\right) k^{\nu}$: $k'_z = \gamma (k_z - v_0 \omega_k),\omega'_{k} = \gamma (\omega_k - v_0 k_z)$, we get
\begin{equation}
	\begin{split}
		&\bra{0} \Phi(\tilde{x}(\tau_2)) \Phi(\tilde{x}(\tau_1))\ket{0} = \frac{1}{(2 \pi R)^2} \times \\
		&\sum_{m=-\infty}^{\infty} \sum_{n=1}^{\infty} \frac{J^2_m(\xi_{mn} \rho_0/R)}{J^2_{\abs{m}+1}(\xi_{mn})} \int_{-\infty}^{\infty} \frac{\dd{k'_z}}{\omega'_k} e^{-i \omega'_k (\tau_2-\tau_1)}.
	\end{split}
\end{equation}
Therefore, contribution of the vacuum fluctuations to the energy shift~[Eq.~\eqref{shiftRF}] for energy level $\ket{b}$ of the inertial atom is
\begin{equation}
	\begin{split}
		&(\delta E_b)_{\text{vf}} = \frac{g^2}{2(2 \pi R)^2} \sum_{d} \abs{\bra{b}\mu(\tau_0)\ket{d}}^2 \sum_{m,n} \frac{J^2_m(\xi_{mn} \rho_0/R)}{J^2_{\abs{m}+1}(\xi_{mn})} \times \\
		& \int_{-\infty}^{\infty} \frac{\dd{k'_z}}{\omega'_k}  \Im\left( \int_{0}^{\infty} \dd{u} e^{-i(\omega'_k - \omega_{bd})u} - \int_{0}^{\infty} \dd{u} e^{-i(\omega'_k + \omega_{bd})u}\right),
	\end{split}
\end{equation}
where Im($ \cdot $) denotes the imaginary part of the argument.
Therefore, the net energy shift due to vacuum fluctuations is obtained as
\begin{equation}\label{LSInertial0}
	\begin{split}
		&\Delta_{\text{vf}} = (\delta E_e)_{\text{vf}} - (\delta E_g)_{\text{vf}}\\
		&= \frac{g^2}{(2 \pi R)^2} \sum_{m=-\infty}^{\infty} \sum_{n=1}^{\infty} \frac{J^2_m(\xi_{mn} \rho_0/R)}{J^2_{\abs{m}+1}(\xi_{mn})} \times\\
		&\int_{-\infty}^{\infty} \frac{\dd{k'_z}}{\omega'_k}  \Im\left( \int_{0}^{\infty} \dd{u} e^{-i(\omega'_k - \omega_0)u} - \int_{0}^{\infty} \dd{u} e^{-i(\omega'_k + \omega_0)u}\right).
	\end{split}
\end{equation}

The contribution of self-reaction to the radiative energy shift is given by Eq.~\eqref{shiftSR}. As $C^{\mathrm{(S,b)}}(u)$ is symmetric under the exchange of $b$ and $d$, it is straightforward to show that contribution of the self-reaction to radiative energy shift in both an inertial and a Rindler two-level atom inside the cylindrical cavity vanishes as it does in free space~\cite{Audretsch1995PRA}. Accordingly, the total energy shift $\Delta_{0}$ is equal to the energy shift due to the vacuum fluctuations, that is, $\Delta_0 = \Delta_{\text{vf}}$. We evaluate the integrals appearing in Eq.~\eqref{LSInertial0} as outlined in Appendix~\ref{apSec: LSinertial} and obtain the radiative energy level shift in an inertial atom to be:
\begin{equation}\label{LSInertial}
\begin{split}
	&\Delta_{0} = \frac{g^2 \omega_0}{\pi^2 (R \omega_0)} \sum_{m = - \tilde{m}}^{\tilde{m}} \sum_{n=1}^{\tilde{n}} \frac{J^2_m(\xi_{mn}\rho_0/R)}{J^2_{\abs{m}+1}(\xi_{mn})} \frac{1}{\xi_{mn}} \times \\
        &\begin{cases}
		\frac{1}{\sqrt{1 - \left(\frac{R\omega_0}{\xi_{mn}}\right)^2}} \left[\tan^{-1}\left(\frac{\sqrt{1 - \left(\frac{R \omega_0}{\xi_{mn}}\right)^2}}{\left(\frac{R \omega_0}{\xi_{mn}}\right)}\right) - \frac{\pi}{2}\right],R\omega_0 < \xi_{mn} \\
		\frac{1}{\sqrt{\left(\frac{R\omega_0}{\xi_{mn}}\right)^2 - 1}} \log\left[\frac{R \omega_0}{\xi_{mn}} + \sqrt{\left(\frac{R\omega_0}{\xi_{mn}}\right)^2 - 1} \right], R \omega_0 \geq \xi_{mn}.
		\end{cases}
  \end{split}
\end{equation}
The variation of $\Delta_{0}/\omega_0 g^2$ as a function of the cavity-detuning parameter $R\omega_0$ is plotted in Fig.~\ref{fig:LSrest}. For the atom's radial position coordinate, we have taken $\rho_0 = 0$ in Eq.~\eqref{LSInertial} and therefore the only non-zero term in the sum over $m$ corresponds to $m=0$. The $\Delta_0$ expression is logarithmically divergent in the sum over $n$. We address the divergence by introducing a UV cutoff $\tilde{n}$ on the sum over $n$ determined by $\xi_{0\tilde{n}}/R = m_{\rm e}/\hbar$, where $m_{\rm e}$ is the rest mass of an electron~\cite{Bethe1947,Milonni1994,Sakurai2006}.
\subsection{Radiative energy level shift in a Rindler atom}\label{Sec: RESnin-results}
The Unruh effect is founded in the fact that the inertial vacuum state of a free field restricted to either of the Rindler wedges is a thermal state at temperature $T = a/2 \pi$, where $a$ is the acceleration of the Rindler observer~\cite{Unruh1976}\footnote{The transition rates of an atom undergoing uniform linear acceleration can show thermal character even if the restricted field state is not thermal. See Refs.~\cite{Rovelli2012,Carballo2019}}. As shown in Appendix~\ref{apSec: UE in CyCav}, in the cavity setup considered here the inertial vacuum of the cavity field restricted to, say, the right Rindler wedge is in fact a thermal state. A uniformly accelerated atom moving along the axis of the cylindrical cavity will therefore experience modified field correlations underlying the Unruh effect. Thus, the radiative energy shift, among other properties, in the atom will be modified with a distinct acceleration-induced signature~(see Appendix~\ref{apSec:acceleration-induced modification} for more details). This acceleration-induced contribution is the signal of interest to us.
In this subsection, we compute the field statistical functions $C^{\mathrm{(R)}}(\tilde{x}(\tau),\tilde{x}(\tau'))$ and $\chi^{\mathrm{(R)}}(\tilde{x}(\tau),\tilde{x}(\tau'))$ for a Rindler atom moving along the axis of the cylindrical cavity. 
 For an atom undergoing uniform linear acceleration along $z$-axis, its trajectory in the lab frame parameterized by its proper time $\tau$ and expressed in cylindrical coordinates is given by~\cite{RindlerRel}
 \begin{align}\label{RindlerTrajec}
 	t(\tau) &= a^{-1} \sinh(a \tau), \nonumber\\
 	z(\tau) &= a^{-1} \cosh(a \tau), \\
 	\rho &= \rho_0, \theta = \theta_0 \nonumber.
 \end{align}

 The symmetric field correlation function \eqref{FieldCorrFxn}, therefore, evaluates  to
 \begin{equation}
 	\begin{split}
 		&C^{\mathrm{(R)}}(\tilde{x}(\tau),\tilde{x}(\tau')) =  \frac{1}{2} \frac{1}{(2 \pi R)^2} \sum_{m=-\infty}^{\infty} \sum_{n=1}^{\infty} \frac{J^2_m(\xi_{mn} \rho_0/R)}{J^2_{\abs{m}+1}(\xi_{mn})} \\
 		& \times \int_{-\infty}^{\infty} \frac{\dd{k'_z}}{\omega'_k} \left( e^{-i (2\omega'_k/a) \sinh(au/2)} +  e^{i (2\omega'_k/a) \sinh(au/2)} \right),
 	\end{split}
 \end{equation}
 where we have defined $ u = \tau - \tau', v = (\tau + \tau')/2$; and have used the expressions $\omega'_k = \omega_k \cosh(a v) - k_z \sinh(a v)$ and $k'_z = k_z \cosh(a v) - \omega_k \sinh(a v)$, for the components of the boosted field momentum $k'^{\mu} = \left(\partial x'^{\mu}/\partial x^{\nu}\right) k^{\nu}$. We have also used $\dd{k_z}/\omega_k = \dd{k'_z}/\omega'_k$. Note that the two-point field correlation function $\expval{\Phi(\tilde{x}(\tau))\Phi(\tilde{x}(\tau'))}{0}$ satisfies the Kubo-Martin-Schwinger (KMS) condition $\expval{\Phi(- u - i \beta)\Phi(0)}{0} =  \expval{\Phi(u) \Phi(0)}{0}$~\cite{Kubo1957,Martin-Schwinger1959,Lidar2019notes}, ensuring that the atom thermalizes in the long-time interaction limit~\cite{Aubry2019}.
 
 The radiative energy shift in an atomic energy eigenstate $\ket{b}$ due to the reservoir (field) fluctuations~[Eq.~\eqref{shiftRF}] is, therefore, given by
 \begin{equation}
 	\begin{split}
 		&(\delta E_b)_{\text{rf}} = - \frac{i g^2}{4(2\pi R)^2} \sum_{d} \abs{\bra{b}\mu(0)\ket{d}}^2 \sum_{m=- \infty}^{\infty} \sum_{n=1}^{\infty}\\ 
 		& \times \frac{J^2_m(\xi_{mn} \rho_0/R)}{J^2_{\abs{m}+1}(\xi_{mn})} \int_{0}^{\infty} \dd{u}  \int_{-\infty}^{\infty} \frac{\dd{k'_z}}{\omega'_k} \left(e^{i \omega_{bd}u} - e^{-i\omega_{bd}u}\right) \\
 		& \times \left( e^{-i (2\omega'_k/a) \sinh(au/2)} +  e^{i (2\omega'_k/a) \sinh(au/2)} \right).
 	\end{split}
 \end{equation}
 For a two-level atom the contribution from reservoir fluctuations to the net radiative shift in the atomic levels is thus given by
 \begin{equation}\label{LSTot}
 	\begin{split}
 		&\tilde{\Delta}_{\text{rf}} = (\delta E_e)_{\text{rf}} - (\delta E_g)_{\text{rf}} \\
        &= \frac{g^2}{(2\pi R)^2} \sum_{m,n} \frac{J^2_m(\xi_{mn} \rho_0/R)}{J^2_{\abs{m}+1}(\xi_{mn})} \int_{0}^{\infty} \dd{u}  \int_{-\infty}^{\infty} \frac{\dd{k'_z}}{\omega'_k} \times \\
 		& \Im\left( e^{-i (2\omega'_k/a) \sinh(au/2)} e^{i \omega_{0} u} +  e^{i (2\omega'_k/a) \sinh(au/2)} e^{i \omega_{0} u} \right).
 	\end{split}
 \end{equation}
\begin{figure}
	\centering
	\includegraphics[width=0.95\linewidth]{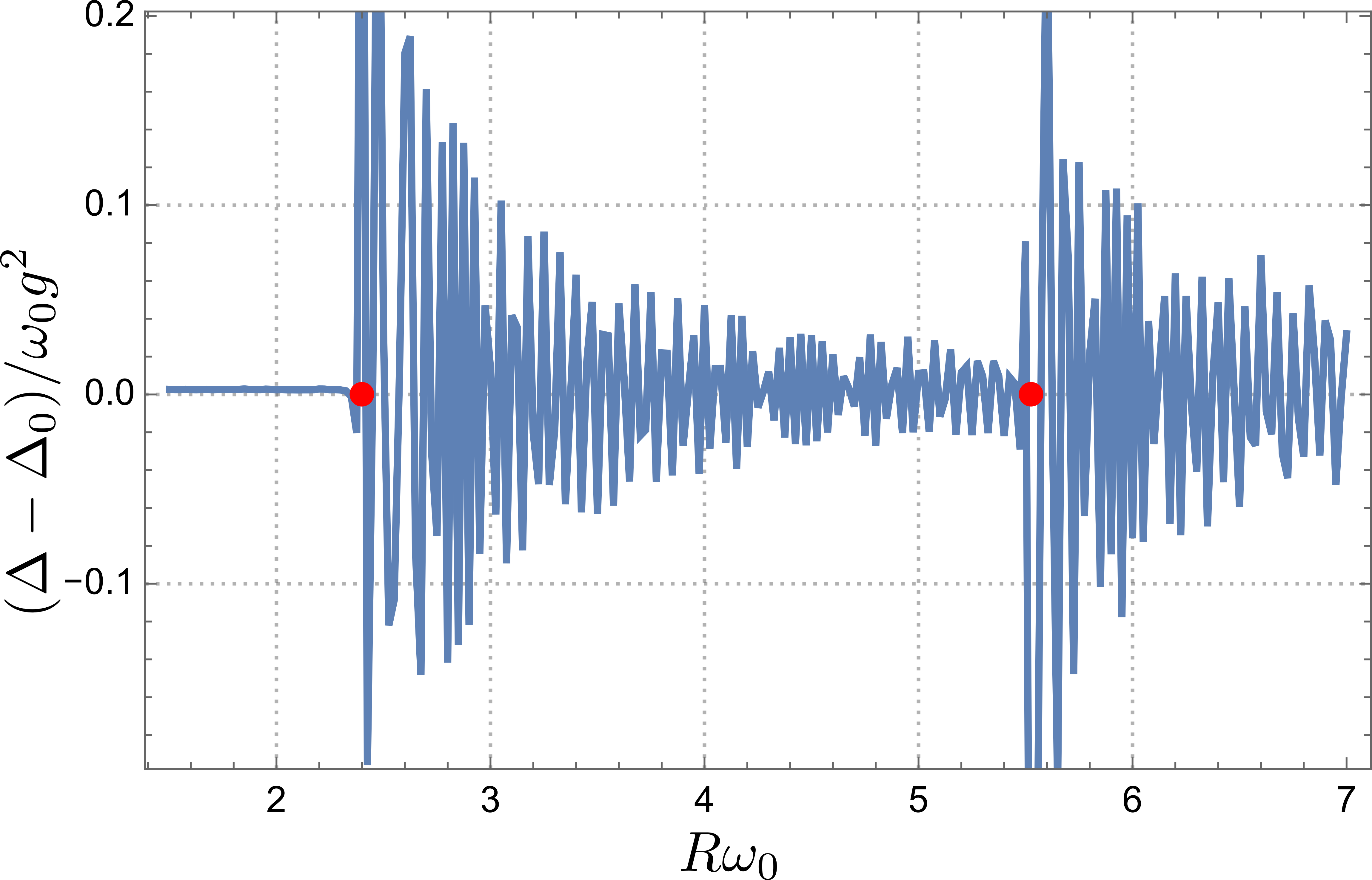}
	\caption{The figure plots the deviation, $\Delta - \Delta_{0}$, of the total energy level shift in a Rindler atom $(\Delta)$ from the radiative shift in an atom moving inertially $(\Delta_{0})$ along the axis of a cylindrical cavity as a function of the cavity-detuning parameter $R \omega_{0}$. The Rindler atom has $\omega_0/a = 10^3$. The red dots placed at $R\omega_0 = \xi_{0n}, n= 1,2$; mark the first two atom-cavity resonance points. Note that the magnitude of $\Delta - \Delta_{0}$ rises in the neighbourhood of the atom-cavity resonance points. In Fig.~\ref{fig:En_vsDetuning}, we investigate $(\Delta - \Delta_0)/\Delta_0$ by zooming in at the first resonance point.}
	\label{fig:alpha103full}
\end{figure}
Since the contribution of self-reaction to the radiative energy shift vanishes, $\tilde{\Delta}_{\text{rf}}$ is, in fact, the total radiative energy shift in the Rindler atom. 
Next, as outlined in Appendix \ref{apSec: integrals}, we cast the expression for the radiative energy shift in a form suitable for obtaining radiative energy shift versus cavity detuning parameter plots. For the total radiative energy level shift of a Rindler atom inside a cylindrical cavity in the $\alpha \equiv \omega_{0}/a \gg 1$ regime we obtain
\begin{widetext}
 \begin{equation}\label{LSTot2}
 	\begin{split}
 			&\Delta \approx \frac{g^2 \omega_0 \alpha}{2 \pi^2 (R\omega_0)^2} \sum_{m = - \tilde{m}}^{\tilde{m}} \sum_{n=1}^{\tilde{n}} \frac{J^2_m(\xi_{mn} \rho_0/R)}{J^2_{\abs{m}+1}(\xi_{mn})} \int_{0}^{\infty} \frac{\dd{\varpi}}{\varpi} \int_{\varpi}^{1} \frac{\dd{\kappa}}{\kappa}  \sin\left(\alpha \ln\frac{\kappa}{\varpi}\right) \cos\Bigg(\frac{\xi_{mn}\alpha}{2R\omega_0}(\varpi - \varpi^{-1} - \kappa + \kappa^{-1})\Bigg)\\
 			&  + \frac{ g^2 \omega_0 \sqrt{\alpha} }{\sqrt{2} \pi  (R \omega_0)^2} \sum_{m=-\tilde{m}}^{\tilde{m}} \sum_{n=1}^{\tilde{n}} \frac{J^2_m(\xi_{mn} \rho_0/R)}{J^2_{\abs{m}+1}(\xi_{mn})} \int_{1}^{\infty}  \frac{\dd{\kappa}}{\kappa}   \sin\left(\alpha \ln(\kappa) - \frac{\xi_{mn}\alpha}{2R\omega_0} (\kappa - \kappa^{-1})\right)  \\
 			&
 			\hspace{9cm} \times \begin{cases}
 				& \frac{(\beta_{mn}^{<} \alpha)^{1/6}}{\left(1 - \left( \frac{\xi_{mn}}{R\omega_0}\right)^2\right)^{1/4}} ~ \text{Ai}[-(\beta_{mn}^{<} \alpha)^{2/3}];~ \frac{\xi_{mn}}{R\omega_0} < 1\\
 				& \frac{\alpha^{1/6}}{3^{2/3} \Gamma(2/3)};~ \frac{\xi_{mn}}{R\omega_0} = 1\\
 				& \frac{(\beta_{mn}^{>} \alpha)^{1/6}}{\left(\left(\frac{\xi_{mn}}{R\omega_0}\right)^2 - 1\right)^{1/4}} ~ \text{Ai}[(\beta_{mn}^{>} \alpha)^{2/3}];~ \frac{\xi_{mn}}{R\omega_0} > 1
 			\end{cases},
 	\end{split}
 \end{equation}
\end{widetext}
where
 \begin{subequations}
 	\begin{equation}
 		\beta^{<}_{mn} \equiv \frac{3}{2} \left(\sech^{-1}\left(\frac{\xi_{mn}}{R \omega_0}\right) - \sqrt{1-\left(\frac{\xi_{mn}}{R \omega_0}\right)^2}\right),
 	\end{equation}
 	
 	\begin{equation}
 		\beta^{>}_{mn} \equiv \frac{3}{2} \left( \sqrt{\left(\frac{\xi_{mn}}{R \omega_0}\right)^2 - 1} - \sec^{-1}\left(\frac{\xi_{mn}}{R \omega_0}\right) \right),
 	\end{equation}
 \end{subequations}
and $\text{Ai}(x)$ is the Airy function. As in the inertial case, for the atom's radial position coordinate we take $\rho_0 = 0$. The radiative energy shift is logarithmically divergent in the sum over $n$. We, therefore, introduce a UV cutoff $\tilde{n}$ on the sum over $n$ determined by $\xi_{0\tilde{n}}/R = m_{\rm e} /\hbar$.
\begin{figure*}
	\centering
	\subfigure[]{
		\includegraphics[width=0.45\linewidth]{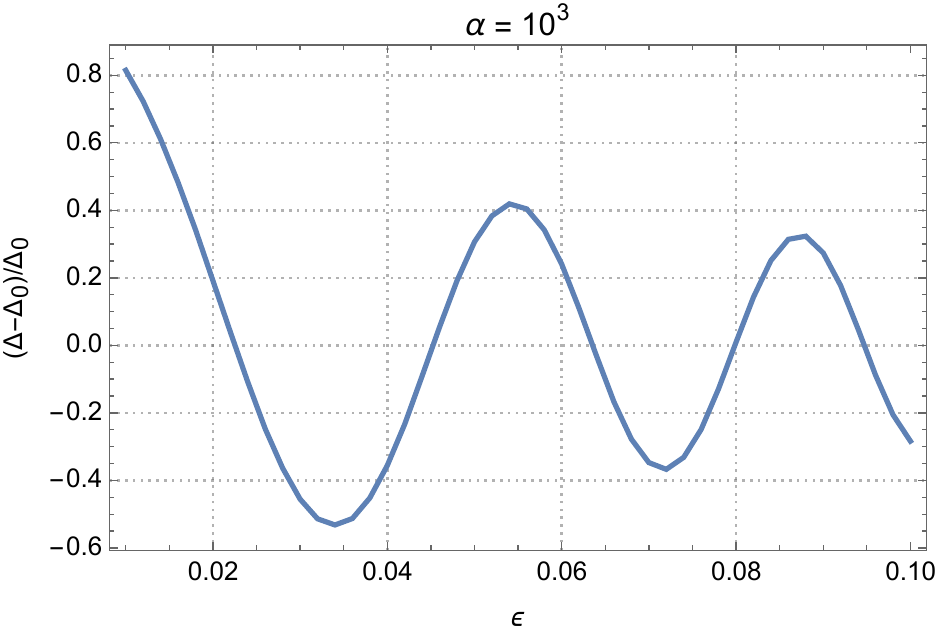} \label{fig:En_vsDetuning10^3}}
	\subfigure[]{
		\includegraphics[width=0.45\linewidth]{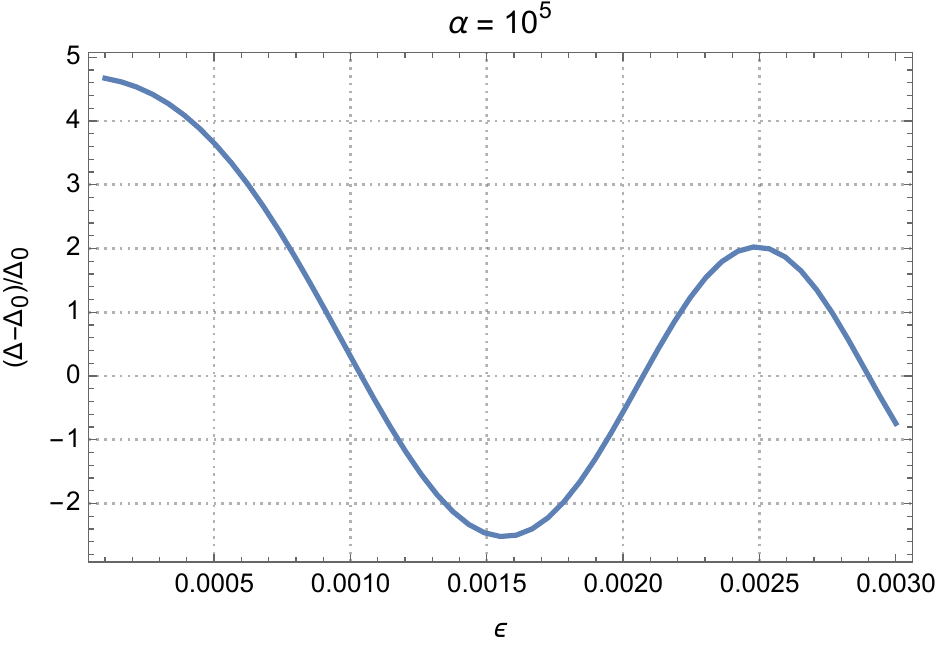} \label{fig:En_vsDetuning10^5}}
	\subfigure[]{
		\includegraphics[width=0.45\linewidth]{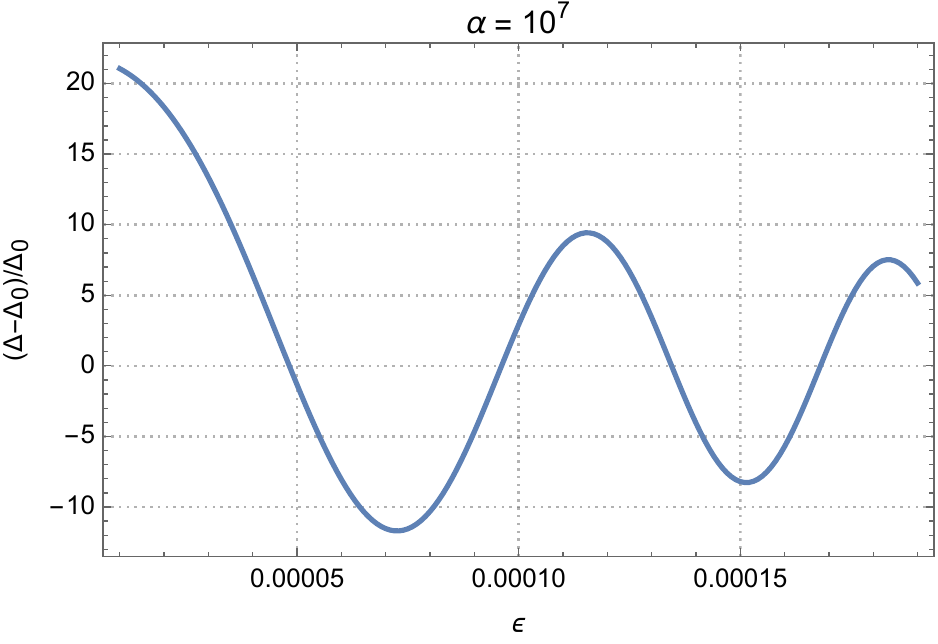} \label{fig:En_vsDetuning10^7}}
	\subfigure[]{
		\includegraphics[width=0.45\linewidth]{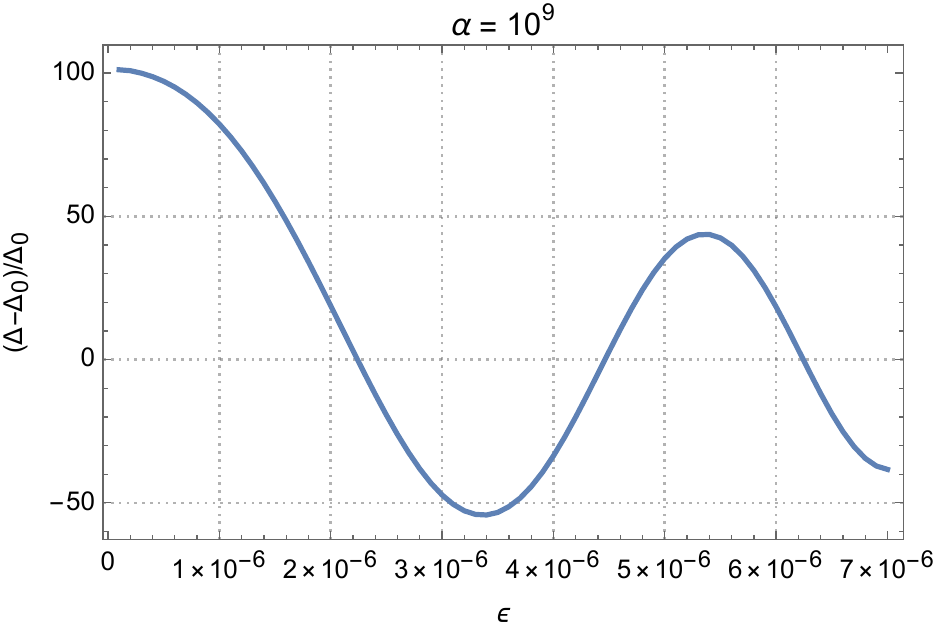} \label{fig:En_vsDetuning10^9}}
	\caption{Relative enhancement of the purely-noninertial energy level shift over the inertial energy level shift, $(\Delta - \Delta_{0})/\Delta_{0}$, as a function of the cavity detuning where we have written $R\omega_0 = \xi_{01} + \epsilon$. We have defined $\alpha \equiv \omega_0/a$.}
	\label{fig:En_vsDetuning}
\end{figure*}

The total radiative energy shift computed above for a Rindler atom has both inertial (i.e., the contribution that survives the $a \to 0$ limit) and purely-noninertial contributions. We can obtain the purely-noninertial contribution as $\Delta - \Delta_0$, where $\Delta_0$ is given by Eq.~\eqref{LSInertial}. The quantity $\Delta - \Delta_{0}$ gives the deviation of the radiative shift in a Rindler atom from that in an inertial atom and, thus, is the signal of interest. The purely-noninertial radiative shift can be expressed as a Cauchy principal value integral over the field mode frequencies $\varpi \equiv \omega_k/\omega_0$ as~(see Appendix~\ref{apSec:acceleration-induced modification}):
\begin{multline}\label{PV-purelynonin1}
     \Delta - \Delta_{0} =\frac{g^2 \omega_0}{(2\pi)^2 (R \omega_0)^2} \sum_{m,n} \frac{J^2_m(\xi_{mn} \rho_0/R)}{J^2_{\abs{m}+1}(\xi_{mn})} \\
     \times \int_{0}^{\infty} \dd{\varpi} f_{mn}(\varpi, \alpha, R\omega_0) \text{P.V.} \left(\frac{1}{\varpi + 1} - \frac{1}{\varpi - 1}\right),
 \end{multline}
 where P.V. denotes Cauchy principal value, and
 \begin{multline*}
     f_{mn}(\varpi, \alpha, R\omega_0) \equiv \frac{4 \alpha}{\pi} \cosh(\pi \varpi \alpha) K^2_{i\varpi \alpha}(\xi_{mn}\alpha/R\omega_0) \\
     - \frac{\Theta\left(\varpi - (\xi_{mn}/R\omega_0)\right)}{\sqrt{\varpi^2 - (\xi_{mn}/R\omega_0)^2}}.
 \end{multline*}
 The magnitude of $\Delta - \Delta_0$ is essentially controlled by the difference in the noninertial and inertial terms captured in $f_{mn}(\varpi, \alpha, R\omega_0)$. For a given atom, one can vary the radius of the cavity looking for a cavity configuration in which either of the two terms dominates the other over a range of field mode frequencies. Given the (principal value) integral in Eq.~\eqref{PV-purelynonin1}, this difference then accumulates over the range of field frequencies, possibly leading to a purely-noninertial signal stronger than the one obtained if atom's emission rate is monitored~\cite{Jaffino2022}.
 Fig.~\ref{fig:alpha103full} shows that the magnitude of $\Delta - \Delta_{0}$ rises abruptly in the neighbourhood of an atom-cavity resonance point. We, therefore, proceed to investigate $\Delta - \Delta_{0}$ more closely in the vicinity of an atom-cavity resonance point.

In Fig.~\ref{fig:En_vsDetuning}, the \textit{relative enhancement}, $\mathcal{F}(\alpha, R \omega_{0}) \equiv (\Delta - \Delta_{0})/\Delta_{0}$, of the purely noninertial energy level shift over the inertial energy level shift, is plotted as a function of the cavity-detuning parameter $R\omega_0$ in the neighbourhood of the first atom-cavity resonance point $\xi_{01}$. Obtaining $\mathcal{F} \gtrsim 1$ in the parameter regime in which the inertial radiative energy shift $\Delta_{0}$ has already been investigated experimentally~(see, for example, ref.~\cite{Marrocco1998}) means that the inertial setup adapted to the noninertial setting considered here can unambiguously resolve the purely-noninertial signal.

From Fig.~\ref{fig:En_vsDetuning} we note that smaller accelerations lead to a larger enhancement but require the specification of the cavity's radius (or equivalently, cavity's normal frequency) with a greater precision. A relative enhancement in the range $1$ to $50$ can be obtained inside a suitably designed cavity. For example, for an atom with $\omega_0 \sim 10$GHz, to obtain a relative enhancement $\sim 50$ with $\alpha \sim 10^9$ (i.e., $a \sim 3 \times 10^{9}~\mathrm{m/s^2}$), the cavity radius $(R_0 \sim 6~\mathrm{cm})$ needs to be specified with an uncertainty no more than $\sim 10$nm.
\begin{table}[!h]
	\begin{center}
		\begin{tabular}{c c c}
            \hline
			$ \alpha = \omega_0c/a $  & $\delta R/R_{0}$ & $\mathcal{F}$  \\
			\hline
            \hline
			$10^5$  & $10^{-5}$ &  $ 1 $\\
			\hline
			$10^7$  & $10^{-6}$ &  $ 10 $\\
			\hline
			$10^9$  & $10^{-7}$ &  $ 50 $\\
			\hline
		\end{tabular}
		\caption{\label{table} The table lists the achievable relative enhancement $\mathcal{F}$ and required relative precision $\delta R/R_{0}$ in the design of the cavity for various values of the Rindler atom's proper acceleration. Here, $R_{0}$ is defined as $\omega_0 R_0/c \equiv \xi_{01}$. In contrast, relative enhancement in emission rate is of the order of 1 for the range of accelerations considered here~\cite{Jaffino2022}.}
	\end{center}
\end{table}
\noindent
We can compare the precision required in the specification of the cavity's radius here with that achieved in an experiment~\cite{Marrocco1998} measuring radiative shift of Rydberg energy levels in an inertial atom between parallel metal plates\textemdash ``interplate distances ranging from $0.5$ to $3$mm were monitored by means of an interferometric measuring device with a maximum uncertainty of $\pm 1 \mu$m~\cite{Marrocco1998}.'' This corresponds to a relative precision $\delta d/d_0 \sim 10^{-3}$ in the determination of the interplate distance. Table~\ref{table} lists achievable relative enhancement of the purely-noninertial signal and correspondingly required relative precision in the cavity design for different values of the Rindler atom's proper acceleration. While the use of a cavity enabled us to make the noninertial response dominant over the inertial one, choice of monitoring the radiative energy shifts in a Rindler atom, in contrast to the transition rates~\cite{Jaffino2022}, allowed us to reap even higher relative enhancement. With Ref.~\cite{Marrocco1998} as a benchmark, Table~\ref{table} testifies to the potential of the undertaken approach for the detection of Unruh effect. It is conceivable that experiments such as in Ref.~\cite{Marrocco1998} can be adapted for measuring radiative energy shifts in linearly accelerated atoms.
\section{Conclusion}\label{Sec: conclusion}
In this work, we have proposed that a suitably modified density of field states, supplemented with a judicious choice of the system property to be monitored, forms an efficient strategy to test the Unruh effect.
At the outset, we mentioned that different system properties might be affected to differing extents due to its noninertial motion. Identifying a system property that not only efficiently captures noninertial effects but also has mature experimental techniques for measurement in a noninertial setup can facilitate a laboratory test of Unruh's prediction.

In this context, we studied the radiative shift in the spectrum of a Rindler atom inside a cylindrical cavity. To illustrate the merit of the approach undertaken, as an example we pointed to an experiment measuring radiative energy shifts in inertial atoms between parallel metal plates which could possibly be adapted to a noninertial setting. 

As known for transition rates, it was shown that even the radiative energy shifts of an accelerating atom receive a significant non inertial component if the cavity size is appropriately designed. Even for marginal accelerations, it was shown that a specific cavity of a precise enough radius will have the energy-level shift essentially entirely controlled by the noninertial contribution.  Thus, this study provides a direct experimental method of accessing the thermality an accelerated atom is expected to perceive via the change in its energy levels. The noninertial effects are argued to be commensurate with (or even larger) than the typical radiative shifts measured in realistic settings.  Given the unprecedented precision with which atomic spectral lines have been determined, the relative enhancement of the purely-noninertial radiative energy shift achieved in this work may facilitate a test of the Unruh effect with current experimental capabilities. This offers one of the most precise and feasible experimental proposals capable of isolating the thermal effect an accelerated frame is supposed to experience in the inertial vacuum. 

Given the precision already attained in measuring the radiative shifts, such a study can also potentially be extended to investigate minute changes in the correlation functions of quantum fields through other dynamical processes or changes in the spacetime curvature, e.g., the presence of a black hole or gravitational waves. Such studies will be taken elsewhere.
\begin{acknowledgments}
	N.A. acknowledges financial support from the University Grants Commission (UGC), Government of India, in the form of a research fellowship (Sr.~No.~2061651285). Research of KL is partially supported by SERB, Government of India, through a MATRICS research grant no.
    MTR/2022/000900. S.K.G. and N.A. acknowledge the financial support from the Interdisciplinary Cyber Physical Systems (ICPS) programme of the Department of Science and Technology, India (Grant No. DST/ICPS/QuST/Theme-1/2019/12). 
\end{acknowledgments}
\onecolumngrid
\appendix
\section{Derivation of radiative energy shift in an inertial atom}\label{apSec: LSinertial}
To evaluate the $k'_z$-integral in Eq.~\eqref{LSInertial0} we make a change of the integration variable to
\begin{equation*}
    \zeta \equiv \frac{\omega'_k + k'_z}{\xi_{mn}/R},
\end{equation*}
and obtain
\begin{equation}
 I=\int_{-\infty}^{\infty} \frac{\dd{k_z}}{\omega_{k}} e^{-i\omega_{k}u}
 =\int_{0}^{\infty} \frac{\dd{\zeta}}{\zeta} \exp\left[-i \frac{\xi_{mn}u}{2 R} \left(\zeta + \zeta^{-1}\right)\right].
\end{equation}
Making use of the integral representation of the Modified Bessel function $K_{\nu}(z)$~\cite{Gradshteyn}:
\begin{equation}
 K_{\nu}(xz)=\frac{z^{\nu}}{2} \int_{0}^{\infty} \frac{\dd t}{t^{\nu+1}} \exp\left[-\frac{x}{2}\left(t+\frac{z^2}{t}\right)\right],
\end{equation}
for $|{\rm arg} (z)|<\pi/4$ or $\abs{\arg{z}} = \pi/4$ and ${\rm Re} (\nu)<1$, we get
\begin{equation}
 I=2K_{0}(i\xi_{mn}u/R).
\end{equation}
Further, using~\cite{Gradshteyn}
\begin{eqnarray}
 \int_{0}^{\infty}\dd{x} e^{-\alpha x} K_{0}(\beta x)
 =\frac{1}{\sqrt{\alpha^2-\beta^2}} \ln\left(\frac{\alpha}{\beta}+\sqrt{\frac{\alpha^2}{\beta^2}-1}\right),
\end{eqnarray}
for $\Re(\alpha+\beta)>0$, to evaluate the $u$-integral, leads to Eq.~\eqref{LSInertial}.

\noindent

\textbf{\textit{Inertial radiative energy shift as a Cauchy Principal Value integral.\textemdash}}
Using~\cite{Heitler1954} 
\begin{equation}
	\lim_{R \rightarrow \infty} \int_{0}^{R} \dd{\kappa} e^{- i \kappa x} = -i\frac{\text{P.V.}}{x} + \pi \delta(x),
\end{equation}
where P.V. denotes Cauchy Principal value; $\dd{k'_z}/\omega'_k = \dd{\omega'_k}/k'_z$; and changing the integration variable to $\varpi \equiv \omega_k/\omega_0$, Eq.~\eqref{LSInertial0} can be expressed as a Cauchy Principal value integral as
\begin{equation}\label{LSInertialPV}
	\begin{split}
		&\Delta_{0} = \frac{g^2 \omega_0}{(2 \pi)^2 (R\omega_0)^2} \sum_{m=-\infty}^{\infty} \sum_{n=1}^{\infty} \frac{J^2_m(\xi_{mn} \rho_0/R)}{J^2_{\abs{m}+1}(\xi_{mn})} \int_{0}^{\infty} \dd{\varpi} \frac{\Theta\left(\varpi - (\xi_{mn}/R\omega_0)\right)}{\sqrt{\varpi^2 - (\xi_{mn}/R\omega_0)^2}} ~\text{P.V.} \left(\frac{1}{\varpi + 1} - \frac{1}{\varpi - 1}\right).
	\end{split}
\end{equation}
Here, $\Theta(x)$ is the Heaviside theta function.

\section{Unruh effect inside a cylindrical cavity}\label{apSec: UE in CyCav}
{\bf \textit{Rindler and Unruh modes of the cavity field.\textemdash}} Solving the Klein-Gordon equation in the Rindler coordinates $(\tau, \xi, \rho, z)$~\cite{RindlerRel} by selecting positive-frequency modes w.r.t. the Killing field $K = \pdv*{}{\tau}$ i.e., $\propto \exp(-i \omega \tau)$, cavity-field modes in the right Rindler wedge are obtained to be
\begin{equation}\label{RRindler-modes2}
	v^{\rm R}_{\omega; mn}(\tau,\xi,\rho,\theta) = \frac{1}{\sqrt{2\pi \omega R^2}} \frac{J_m\left(\xi_{mn} \rho/R\right)}{J_{\abs{m}+1}(\xi_{mn})} \left[\frac{2 \omega \sinh(\pi \omega/a)}{\pi^2 a}\right]^{1/2}  e^{im\theta} K_{i\omega/a}\left(\frac{\xi_{mn}}{Ra} e^{a\xi}\right) e^{-i\omega\tau},
\end{equation}
where $m \in \mathbb{Z}, 1 \leq n < \infty$. The positive-frequency modes in the left Rindler wedge $v^{\rm L}_{\omega; mn}(\bar{\tau},\bar{\xi},\rho,\theta)$ are obtained from $v^{\rm R}_{\omega; mn}(\tau,\xi,\rho,\theta)$ by replacing $\tau$ and $\xi$ by $\bar{\tau}$ and $\bar{\xi}$, respectively~\cite{Crispino2008}. The expansion of the field in the left and right Rindler wedges is, therefore, given by
\begin{multline}\label{RindlerPhi}
	\Phi(\tau, \xi, \rho, \theta) = \sum_{m=-\infty}^{\infty} \sum_{n=1}^{\infty} \int_{0}^{\infty} \dd{\omega} \Big(a^{\rm R}_{\omega; mn} v^{\rm R}_{\omega; mn}(\tau,\xi,\rho,\theta) + a^{\rm R^{\dagger}}_{\omega; mn} v^{\rm R^{*}}_{\omega; mn}(\tau,\xi,\rho,\theta) \\
	+ a^{\rm L}_{\omega; mn} v^{\rm L}_{\omega; mn}(\bar{\tau},\bar{\xi},\rho,\theta) + a^{\rm L^{\dagger}}_{\omega; mn} v^{\rm L^{*}}_{\omega; mn}(\bar{\tau},\bar{\xi},\rho,\theta)\Big),
\end{multline}
with the Rindler vacuum state $\ket{0_{\rm R}}$ defined as $a^{\rm R}_{\omega; mn} \ket{0_{\rm R}} = a^{\rm L}_{\omega; mn} \ket{0_{\rm R}} = 0$, for all $\{\omega,m,n\}$. We can expand the Rindler modes in terms of the inertial modes $f_{\omega_k;mn}$ as
\begin{subequations}
	\begin{equation}
		v^{\rm R}_{\omega;mn} = \sum_{m,n} \int_{-\infty}^{\infty} \dd{k}_z \left(\alpha^R_{\omega k_z; mn} f_{\omega_k;mn} + \beta^R_{\omega k_z; mn} f^*_{\omega_k;mn}\right),
	\end{equation}
	\begin{equation}
		v^{\rm L}_{\omega;mn} = \sum_{m,n} \int_{-\infty}^{\infty} \dd{k}_z \left(\alpha^L_{\omega k_z; mn} f_{\omega_k;mn} + \beta^L_{\omega k_z; mn} f^*_{\omega_k;mn}\right),
	\end{equation}
\end{subequations}
and establish that the Bogoliubov coefficients $\alpha_{\omega k_z; mn}$ and $\beta_{\omega k_z; mn}$ satisfy the following relations
	\begin{subequations}\label{Bog-Relations}
	\begin{equation}\label{Bog-Relations1}
		\alpha^{\rm L}_{\omega k_z;mn} = \alpha^{\rm R}_{\omega, -k_z;mn}, ~ \beta^{\rm L}_{\omega k_z;mn} = \beta^{\rm R}_{\omega, -k_z;mn};
	\end{equation}
	\begin{equation}\label{Bog-Relations2}
		e^{-\pi \omega/a}\alpha^{\rm R}_{\omega k_z; mn} = - \beta^{\rm R}_{\omega k_z; mn}, ~ \alpha^{\rm R^*}_{\omega k_z; mn} = \alpha^{\rm R}_{\omega, -k_z; mn}, ~ \beta^{\rm R^*}_{\omega, -k_z; mn} =  \beta^{\rm R}_{\omega,k_z; mn}.
	\end{equation}
\end{subequations}

With Eqs.~\eqref{Bog-Relations}, using $v^{\rm R}_{\omega; mn}$ and $v^{\rm L}_{\omega; mn}$, we can construct modes, known as the Unruh modes, which are linear combinations of purely positive frequency modes $e^{-i\omega_k t}$ in the Minkowski spacetime. The modes defined as
\begin{align}\label{w+-}
	w_{-\omega; mn} &\equiv \frac{v^{\rm R}_{\omega; mn} + e^{-\pi\omega/a} v^{\rm L*}_{\omega; -mn}}{\sqrt{1 - e^{- 2 \pi \omega/a}}} \\
	w_{+\omega; mn} &\equiv \frac{v^{\rm L}_{\omega; mn} + e^{-\pi\omega/a} v^{\rm R*}_{\omega; -mn}}{\sqrt{1 - e^{- 2 \pi \omega/a}}}
\end{align}
are linear combinations of purely positive frequency modes $e^{-i\omega_k t}$ in the Minkowski spacetime.
The field can be expanded and quantized in terms of the Unruh modes as:
\begin{equation}\label{Phi-w}
	\Phi(t,z,\vb{x}_{\perp}) \equiv \sum_{n=1}^{\infty} \sum_{m=-\infty}^{\infty} \int_{0}^{\infty} \dd{\omega} \Big[w_{-\omega; mn} b_{-\omega; mn} + w^{}_{+\omega; mn} b^{}_{+\omega; mn} + w^*_{-\omega; mn} b^{\dagger}_{-\omega; mn} + w^{*}_{+\omega; mn} b^{\dagger}_{+\omega; mn}\Big],
\end{equation}
where
\begin{subequations}\label{Unruh-ops1}
	\begin{align}
		b_{-\omega; mn} &\equiv \frac{a^R_{\omega; mn} - e^{-\pi\omega/a} a^{L^{\dagger}}_{\omega;-mn}}{\sqrt{1 - e^{- 2 \pi \omega/a}}},~\mathrm{and} \\
		b_{+\omega; mn} &\equiv \frac{a^L_{\omega; mn} - e^{-\pi\omega/a} a^{R^{\dagger}}_{\omega;-mn}}{\sqrt{1 - e^{- 2 \pi \omega/a}}}. 
	\end{align}
\end{subequations}
Also note that the Unruh modes satisfy
\begin{subequations}\label{unruhmode-orthonormality}
	\begin{align}
		\left(w_{\pm \omega; mn}, w_{\pm \omega'; m'n'}\right) &= \delta(\omega - \omega') \delta_{m,m'} \delta_{n,n'}, \\
		\left(w^{*}_{\pm \omega; mn}, w^{*}_{\pm \omega'; m'n'}\right) &= - \delta(\omega - \omega') \delta_{m,m'} \delta_{n,n'}.
	\end{align}
\end{subequations}
Since $w_{\mp \omega; mn}$ are purely positive frequency modes in the Minkowski spacetime, from Eq.~\eqref{Phi-w} we have
\begin{equation}\label{Unruh-ops2}
	b_{-\omega; mn} \ket{0} = 0, b_{+\omega; mn} \ket{0} = 0,
\end{equation}
for all $\{\omega,m,n\}$.
In Sec~\eqref{Sec: RESnin-results}, we have used the field quantization~\eqref{Phi-w} in terms of the Unruh modes to compute the radiative energy shifts in a Rindler atom~[see Eq.~\eqref{RES-UnruhModes}].

\textbf{\textit{Inertial vacuum state of the cavity field restricted to a Rindler wedge.\textemdash}}
Using Eqs.~\eqref{Unruh-ops1}, following ref.~\cite{Crispino2008} one can show that the inertial vacuum of the cavity field restricted to, say, the right Rindler wedge is in fact a thermal state given by
\begin{equation}
	\rho_{\rm R} = \bigotimes_{\{\omega_i; m,n\}} \sum_{N_{\omega_i;mn} = 0}^{\infty} (1 - e^{-2 \pi \omega_i/a}) e^{- 2 \pi N_{\omega_i;mn}  \omega_i/a} \ket{N_{\omega_i;mn}, {\rm R}} \bra{N_{\omega_i;mn}, {\rm R}},
\end{equation}
hence establishing the Unruh effect inside the cylindrical cavity. Here, $\omega_i$ is the frequency of a Rindler mode, $N_{\omega_i;mn}$ is the average number of excitations in a Rindler mode $\{\omega_i; m,n\}$, and $\ket{N_{\omega_i;mn}, {\rm R}} $ denotes right Rindler mode $\{\omega_i; m,n\}$ populated by $N$ excitations.
\section{Acceleration-induced modification of the radiative energy shift}\label{apSec:acceleration-induced modification}
To manifest the acceleration-induced modification of the radiative energy shift, we first use the field expansion in terms of the Unruh modes. Proceeding as we did for the inertial atom, the radiative energy level shift in a Rindler atom moving along the axis of the cylindrical cavity is obtained as
 \begin{multline}\label{RES-UnruhModes}
 		\Delta =\frac{g^2 \omega_0}{(2\pi)^2 (R \omega_0)^2}  \frac{4 \alpha}{\pi} \sum_{m,n} \frac{J^2_m(\xi_{mn} \rho_0/R)}{J^2_{\abs{m}+1}(\xi_{mn})} \int_{0}^{\infty} \dd{\varpi} \cosh(\pi \varpi \alpha) K^2_{i\varpi \alpha}(\xi_{mn}\alpha/R\omega_0) ~\text{P.V.} \left(\frac{1}{\varpi + 1} - \frac{1}{\varpi - 1}\right),
 \end{multline}
 where $\alpha \equiv \omega_0/a$.
 Here, as for the inertial case, we have used the fact that the self-reaction does not contribute to the radiative energy shift in a two-level Rindler atom. Therefore, the total radiative energy shift is $\Delta = \Delta_{\mathrm{vf}}$.

 From Eqs.~\eqref{LSInertialPV} and \eqref{RES-UnruhModes} we can obtain the purely-noninertial contribution (i.e., the acceleration-induced modification) to the radiative energy shift as
 \begin{multline}\label{PV-purelynonin}
     \Delta - \Delta_{0} =\frac{g^2 \omega_0}{(2\pi)^2 (R \omega_0)^2} \sum_{m,n} \frac{J^2_m(\xi_{mn} \rho_0/R)}{J^2_{\abs{m}+1}(\xi_{mn})} \int_{0}^{\infty} \dd{\varpi}
 		 \Bigg[\frac{4 \alpha}{\pi} \cosh(\pi \varpi \alpha) K^2_{i\varpi \alpha}(\xi_{mn}\alpha/R\omega_0) - \frac{\Theta\left(\varpi - (\xi_{mn}/R\omega_0)\right)}{\sqrt{\varpi^2 - (\xi_{mn}/R\omega_0)^2}}\Bigg] \\
    \times \text{P.V.} \left(\frac{1}{\varpi + 1} - \frac{1}{\varpi - 1}\right).
 \end{multline}
The acceleration-induced modification is thus dictated by the term within square brackets in the above expression.

 \section{Derivation of radiative energy shift in the Rindler atom}\label{apSec: integrals}
Let us write Eq.~\eqref{LSTot} as
 \begin{equation}
		\tilde{\Delta}_{\text{rf}} \equiv \frac{g^2}{(2\pi R)^2} \sum_{m,n} \frac{J^2_m(\xi_{mn} \rho_0/R)}{J^2_{\abs{m}+1}(\xi_{mn})} \Im\left(\mathcal{I}^{(1)}_{mn} + \mathcal{I}^{(2)}_{mn}\right),
\end{equation}
and first consider
\begin{equation}
	\mathcal{I}^{(1)}_{mn} \equiv \int_{0}^{\infty} \dd{u}  \int_{-\infty}^{\infty} \frac{\dd{k'_z}}{\omega'_k} e^{i \omega_{0} u} e^{i (2\omega'_k/a) \sinh(au/2)}.
\end{equation}
Effecting a change of variables $\left\{k'_z,u\right\} \mapsto \left\{\zeta,\lambda\right\}$, with
	\begin{equation}
		k'_z \mapsto \zeta \equiv \frac{\omega'_k + k'_z}{\xi_{mn}/R},~
		u \mapsto \lambda \equiv \exp(au/2),
	\end{equation}
we obtain
\begin{equation}
		\mathcal{I}^{(1)}_{mn} = \frac{2}{a} \int_{1}^{\infty} \dd{\lambda}  \int_{0}^{\infty}  \frac{\dd{\zeta}}{\zeta} \lambda^{(2i\omega_{0}/a-1)}  e^{i (\xi_{mn}/R) (\zeta + \zeta^{-1}) (\lambda - \lambda^{-1})/(2a)}.
\end{equation}

We already have $\mathcal{I}^{(1)}_{mn}$ in a form suitable for obtaining energy level shift versus $R\omega_0$ plots, but we further massage the integrals and show that at least one part of the integrals can be evaluated in closed form in terms of modified Bessel-K and incomplete Bessel-K functions. To this end, we effect a second change of variables: $\left\{\zeta,\lambda\right\} \mapsto \left\{\varpi,\kappa\right\}$ with $\kappa = \zeta \lambda$ and $\varpi = \zeta/\lambda$. The Jacobian for this transformation is $ 1/2\varpi$, therefore, with $\dd{\zeta} \dd{\lambda} = (1/2\varpi)\dd{\varpi}\dd{\kappa}$, we obtain
\begin{equation}
		\mathcal{I}^{(1)}_{mn} = \frac{1}{a} \int_{0}^{\infty} \frac{\dd{\varpi}}{\varpi^{(i\alpha + 1)}} e^{-i (\xi_{mn}/R\omega_0) ( \varpi - \varpi^{-1})\alpha/2}  \int_{\varpi}^{\infty} \frac{\dd{\kappa}}{\kappa^{(-i\alpha + 1)}} e^{i (\xi_{mn}/R\omega_0) (\kappa - \kappa^{-1})\alpha/2},
\end{equation}
where we have defined $\alpha = \omega_0/a$.
Similarly, we can obtain
\begin{equation}
		\mathcal{I}^{(2)}_{mn} =  \frac{1}{a} \int_{0}^{\infty} \frac{\dd{\varpi}}{\varpi^{(i\alpha + 1)}}  e^{i (\xi_{mn}/R\omega_0) ( \varpi - \varpi^{-1})\alpha/2}  \int_{\varpi}^{\infty}  \frac{\dd{\kappa}}{\kappa^{(-i\alpha + 1)}}  e^{-i (\xi_{mn}/R\omega_0) (\kappa - \kappa^{-1})\alpha/2}.
\end{equation}
Next, for brevity we define
\begin{subequations}
	\begin{equation}
		f(\varpi) \equiv \frac{1}{\varpi^{(i\alpha + 1)}}  e^{-i (\xi_{mn}/R\omega_0) ( \varpi - \varpi^{-1})\alpha/2}, 
	\end{equation}
	\begin{equation}
		g(\varpi) \equiv \frac{1}{\varpi^{(i\alpha + 1)}}  e^{i (\xi_{mn}/R\omega_0) ( \varpi - \varpi^{-1})\alpha/2}, 
	\end{equation}
\end{subequations}
and write
\begin{equation}
		\mathcal{I}^{(1)}_{mn} = \frac{1}{a} \int_{0}^{\infty} \dd{\varpi} \int_{\varpi}^{1}  \dd{\kappa} f(\varpi) f^*(\kappa) + \frac{1}{a} \int_{0}^{\infty} \dd{\varpi}  \int_{1}^{\infty} \dd{\kappa} f(\varpi) f^*(\kappa),
\end{equation}
and
\begin{equation}
		\mathcal{I}^{(2)}_{mn} = \frac{1}{a} \int_{0}^{\infty} \dd{\varpi} \int_{\varpi}^{1}  \dd{\kappa} g(\varpi) g^*(\kappa) + \frac{1}{a} \int_{0}^{\infty} \dd{\varpi} g(\varpi)  \int_{1}^{\infty} \dd{\kappa} g^*(\kappa).
\end{equation}
Therefore,
\begin{equation}
		\Im{\mathcal{I}^{(1)}_{mn} + \mathcal{I}^{(2)}_{mn}} \\
		= \frac{1}{a} \Bigg[\int_{0}^{\infty} \dd{\varpi} \int_{\varpi}^{1}  \dd{\kappa} \Im\left\{f(\varpi) f^*(\kappa) + g(\varpi) g^*(\kappa)\right\} 
		+  \int_{0}^{\infty} \dd{\varpi}  \int_{1}^{\infty} \dd{\kappa} \Im\left\{f(\varpi) f^*(\kappa) + g(\varpi) g^*(\kappa)\right\} \Bigg].
\end{equation}
Note that
\begin{equation}\label{integrand1}
	\begin{split}
		\Im \left( f(\varpi) f^*(\kappa) + g(\varpi) g^*(\kappa) \right) &= 2 \Im\left(\frac{\kappa^{(i\alpha - 1)}}{\varpi^{(i\alpha + 1)}} \cos\Bigg\{\frac{\xi_{mn}\alpha}{2R\omega_0}(\varpi - \varpi^{-1} - \kappa + \kappa^{-1})\Bigg\} \right) \\
		&= \frac{2}{\varpi \kappa}  \sin\left(\alpha \ln\frac{\kappa}{\varpi}\right) \cos\Bigg\{\frac{\xi_{mn}\alpha}{2R\omega_0}(\varpi - \varpi^{-1} - \kappa + \kappa^{-1})\Bigg\}.
	\end{split}
\end{equation}
Before we proceed further, some comments about $\Im \left( f(\varpi) f^*(\kappa) + g(\varpi) g^*(\kappa) \right)$ are in order:
\begin{enumerate}
	\item \label{It1}  We are interested in the behavior of the radiative shift of the atom's spectra in the regime such that
	\begin{enumerate}
		\item $\omega_0/a \equiv \alpha \gg 1$ i.e., the ``low'' acceleration regime, and
		\item the detuning parameter is such that $R\omega_0 = \xi_{01} + \epsilon$, where $\epsilon \ll 1$,
	\end{enumerate}
	therefore we have either $\xi_{0n}/R\omega_0 \lesssim 1$ or $\xi_{0n}/R\omega_0 > 1$.
	
	\item Referring to Eq.~\eqref{integrand1}, Statement \eqref{It1} means that the integrand in $\Im(\mathcal{I}^{(1)}_{mn} + \mathcal{I}^{(2)}_{mn})$ is a relatively slowly varying function $1/\varpi \kappa$ multiplied by a very rapidly oscillating function for sufficiently small or large values of $\varpi$ or $\kappa$, unless $\varpi \approx \kappa$, in which case $(\varpi - \varpi^{-1} - \kappa + \kappa^{-1}) \approx 0$ and $\ln(\kappa/\varpi) \approx 0$. But note that when $\varpi \approx \kappa$, $\Im \left( f(\varpi) f^*(\kappa) + g(\varpi) g^*(\kappa) \right)$ is itself very small, i.e., $\approx 0$. 
	
Further, there is an overall suppression by $1/\varpi \kappa$ for large values of $\varpi$ and $\kappa$.
\end{enumerate}
We use these observations for numerical evaluation of the integrals.
Further, using $\int_0^{\infty} \dd{y} y^{-1-\nu} e^{\mp 2i\omega (y - y^{-1})} = 2 e^{\pm  i \pi \nu /2} K_{\nu}(4\omega)$, we have $\int_{0}^{\infty} \dd{\varpi} f(\varpi) = 2 e^{-\pi \omega_0/(2a)}K_{i\omega_0/a}(\xi_{mn}/Ra)$, which leads to
\begin{equation}
		\mathcal{I}^{(1)}_{mn} = \frac{1}{a} \int_{0}^{\infty} \dd{\varpi} \int_{\varpi}^{1}  \dd{\kappa} f(\varpi) f^*(\kappa) \\
		+ \frac{2}{a} e^{-\pi \omega_0/(2a)}K_{i\omega_0/a}(\xi_{mn}/Ra) \int_{1}^{\infty} \dd{\kappa} f^*(\kappa),
\end{equation}

and using $\int_{0}^{\infty} \dd{\varpi} g(\varpi) = 2 e^{\pi\omega_0/(2a)}K_{i\omega_0/a}(\xi_{mn}/Ra)$ we have
\begin{equation}
		\mathcal{I}^{(2)}_{mn} = \frac{1}{a} \int_{0}^{\infty} \dd{\varpi} \int_{\varpi}^{1}  \dd{\kappa} g(\varpi) g^*(\kappa) + \frac{2}{a} e^{\pi \omega_0/(2a)}K_{i\omega_0/a}(\xi_{mn}/Ra)  \int_{1}^{\infty} \dd{\kappa} g^*(\kappa).
\end{equation}
The integrals $\int_{1}^{\infty} \dd{\kappa} f(\kappa)$ and $\int_{1}^{\infty} \dd{\kappa} g(\kappa)$ are known as the Incomplete Bessel-K function which is defined as~\cite{Harris2008}:
\begin{equation}
	K_{\nu}(x,y) = \int_{1}^{\infty} \frac{\dd{t}}{t^{\nu + 1}} e^{- x t - y/t};
\end{equation}
we, however, will maintain them in their integral form only.
Using
\begin{subequations}
	\begin{align*}
		\Im(g^*(\kappa)) &= \Im\left( \kappa^{i\alpha - 1} e^{- i (\xi_{mn}/R) (\kappa - \kappa^{-1})/(2a)} \right) = \kappa^{-1} \sin\left[\alpha \ln{\kappa} - (\xi_{mn}/R) (\kappa - \kappa^{-1})/(2a) \right],~\text{and}\\
		\Im(f^*(\kappa)) &= \Im\left( \kappa^{i\alpha - 1} e^{ i (\xi_{mn}/R) (\kappa - \kappa^{-1})/(2a)} \right) = \kappa^{-1} \sin\left[(\xi_{mn}/R) (\kappa - \kappa^{-1})/(2a) + \alpha \ln{\kappa} \right],
	\end{align*}
\end{subequations}
we finally obtain
\begin{equation}
	\begin{split}
		\Im{\mathcal{I}^{(1)}_{mn} + \mathcal{I}^{(2)}_{mn}} &= \frac{2}{a} \Bigg[ \int_{0}^{\infty} \dd{\varpi} \int_{\varpi}^{1}  \dd{\kappa} \frac{1}{\varpi \kappa}  \sin\left(\alpha \ln\frac{\kappa}{\varpi}\right) \cos\Bigg\{\frac{\xi_{mn}\alpha}{2R\omega_0}(\varpi - \varpi^{-1} - \kappa + \kappa^{-1})\Bigg\} + 2 e^{-\pi \omega_0/(2a)}K_{i\omega_0/a}\left(\frac{\xi_{mn}}{Ra}\right) \\
		& \times \int_{1}^{\infty} \dd{\kappa} \kappa^{-1} \sin\left[(\xi_{mn}/R) (\kappa - \kappa^{-1})/(2a) + \alpha \ln{\kappa} \right] + 2 e^{\pi \omega_0/(2a)}K_{i\omega_0/a}(\xi_{mn}/Ra)  \\
		& \times\int_{1}^{\infty} \dd{\kappa} \kappa^{-1} \sin\left[\alpha \ln{\kappa} - (\xi_{mn}/R) (\kappa - \kappa^{-1})/(2a) \right]   \Bigg],
	\end{split}
\end{equation}
To cast the integrals in a form suitable for obtaining the radiative energy level shift versus cavity-detuning parameter $R\omega_0$ plots. To this end, we effect two successive changes of variables:
 \begin{subequations}\label{1stChange}
 	\begin{align}
 		k'_z &\mapsto \zeta \equiv \frac{\omega'_k + k'_z}{\xi_{mn}/R},\\ 
 		u &\mapsto \lambda \equiv \exp(au/2),
 	\end{align}
 \end{subequations}
followed by
$\left\{\zeta,\lambda\right\} \mapsto \left\{\varpi,\kappa\right\}$ with $\kappa = \zeta \lambda$ and $\varpi = \zeta/\lambda$~\cite{Crispino2008}. This allows us to cast the radiative energy level shift in the following form:
\begin{equation}\label{LSTot1}
		\begin{split}
			&\Delta_{\text{rf}} = \frac{g^2 \omega_0 \alpha}{2 \pi^2 (R\omega_0)^2} \sum_{m=-\infty}^{\infty} \sum_{n=1}^{\infty} \frac{J^2_m(\xi_{mn} \rho_0/R)}{J^2_{\abs{m}+1}(\xi_{mn})} \Bigg[ \int_{0}^{\infty} \dd{\varpi} \int_{\varpi}^{1}  \dd{\kappa} \frac{1}{\varpi \kappa}  \sin\left(\alpha \ln\frac{\kappa}{\varpi}\right) \cos\Bigg\{\frac{\xi_{mn}\alpha}{2R\omega_0}(\varpi - \varpi^{-1} - \kappa + \kappa^{-1})\Bigg\} \\
			& + K_{i\alpha}(\xi_{mn}\alpha/R\omega_0) \Bigg\{e^{-\pi \alpha/2} \int_{1}^{\infty} \dd{\kappa} \kappa^{-1} \sin\left[(\xi_{mn}\alpha/R\omega_0) (\kappa - \kappa^{-1})/2 + \alpha \ln{\kappa} \right] \\
			&\hspace{7cm} + e^{\pi \alpha/2} \int_{1}^{\infty} \dd{\kappa} \kappa^{-1} \sin\left[\alpha \ln{\kappa} - (\xi_{mn}\alpha/R\omega_0) (\kappa - \kappa^{-1})/2 \right] \Bigg\}  \Bigg],
		\end{split}
\end{equation}
where $K_{\nu}(x)$ is the modified Bessel function of the second kind. Note that the second term in Eq.~\eqref{LSTot1} is exponentially suppressed. Therefore, for simplification, we can neglect this term in the $\alpha \gg 1$ regime. Also note that since the contribution of self-reaction to the radiative energy shift vanishes, $\tilde{\Delta}_{\text{vf}}$ is, in fact, the total radiative energy shift in the Rindler atom. 

For the trajectory of the Rindler atom, we assume $\rho_0 = 0$. Therefore, the only surviving term from the sum over $m$ would be $m=0$. Equation \eqref{LSTot1} can be further simplified using the expansion of $K_{i\nu}(z \nu)$ for $\nu \gg 1, \nu \in \mathbb{R}, \abs{\arg z} < \pi$~\cite{Olver1997}:
 \begin{equation}\label{asymK2}
 	K_{i \nu}(x \nu) \rightarrow \frac{e^{-\pi \nu /2} \pi \sqrt{2}}{\sqrt{\nu}} \begin{cases}
 		& \frac{(\beta^{<} \nu)^{1/6}}{(1 - x^2)^{1/4}} ~ \text{Ai}[-(\beta^{<} \nu)^{2/3}];~ x < 1\\
 		& \frac{\nu^{1/6}}{3^{2/3} \Gamma(2/3)};~ x = 1\\
 		& \frac{(\beta^{>} \nu)^{1/6}}{(x^2 - 1)^{1/4}} ~ \text{Ai}[(\beta^{>} \nu)^{2/3}];~ x > 1
 	\end{cases}
 \end{equation}
 where
 	\begin{equation}
 		\beta^{<} \equiv \frac{3}{2} \left(\sech^{-1}x - \sqrt{1-x^2}\right),~~ \beta^{>} \equiv \frac{3}{2} \left( \sqrt{x^2 - 1} - \sec^{-1}x \right),
 	\end{equation}
and $\text{Ai}(x)$ is the Airy function. Using the asymptotic approximation for $K_{i\nu}(x\nu)$ in Eq.~\eqref{LSTot1} leads to Eq.~\eqref{LSTot2} of the main text.
\newpage
\twocolumngrid


\begin{thebibliography}{62}%
		\makeatletter
		\providecommand \@ifxundefined [1]{%
			\@ifx{#1\undefined}
		}%
		\providecommand \@ifnum [1]{%
			\ifnum #1\expandafter \@firstoftwo
			\else \expandafter \@secondoftwo
			\fi
		}%
		\providecommand \@ifx [1]{%
			\ifx #1\expandafter \@firstoftwo
			\else \expandafter \@secondoftwo
			\fi
		}%
		\providecommand \natexlab [1]{#1}%
		\providecommand \enquote  [1]{``#1''}%
		\providecommand \bibnamefont  [1]{#1}%
		\providecommand \bibfnamefont [1]{#1}%
		\providecommand \citenamefont [1]{#1}%
		\providecommand \href@noop [0]{\@secondoftwo}%
		\providecommand \href [0]{\begingroup \@sanitize@url \@href}%
		\providecommand \@href[1]{\@@startlink{#1}\@@href}%
		\providecommand \@@href[1]{\endgroup#1\@@endlink}%
		\providecommand \@sanitize@url [0]{\catcode `\\12\catcode `\$12\catcode `\&12\catcode `\#12\catcode `\^12\catcode `\_12\catcode `\%12\relax}%
		\providecommand \@@startlink[1]{}%
		\providecommand \@@endlink[0]{}%
		\providecommand \url  [0]{\begingroup\@sanitize@url \@url }%
		\providecommand \@url [1]{\endgroup\@href {#1}{\urlprefix }}%
		\providecommand \urlprefix  [0]{URL }%
		\providecommand \Eprint [0]{\href }%
		\providecommand \doibase [0]{https://doi.org/}%
		\providecommand \selectlanguage [0]{\@gobble}%
		\providecommand \bibinfo  [0]{\@secondoftwo}%
		\providecommand \bibfield  [0]{\@secondoftwo}%
		\providecommand \translation [1]{[#1]}%
		\providecommand \BibitemOpen [0]{}%
		\providecommand \bibitemStop [0]{}%
		\providecommand \bibitemNoStop [0]{.\EOS\space}%
		\providecommand \EOS [0]{\spacefactor3000\relax}%
		\providecommand \BibitemShut  [1]{\csname bibitem#1\endcsname}%
		\let\auto@bib@innerbib\@empty
		\bibitem [{\citenamefont {Fulling}(1973)}]{Fulling1973}%
		\BibitemOpen
		\bibfield  {author} {\bibinfo {author} {\bibfnamefont {S.~A.}\ \bibnamefont {Fulling}},\ }\bibfield  {title} {\bibinfo {title} {Nonuniqueness of canonical field quantization in {R}iemannian space-time},\ }\href {https://doi.org/10.1103/PhysRevD.7.2850} {\bibfield  {journal} {\bibinfo  {journal} {Phys. Rev. D}\ }\textbf {\bibinfo {volume} {7}},\ \bibinfo {pages} {2850} (\bibinfo {year} {1973})}\BibitemShut {NoStop}%
		\bibitem [{\citenamefont {Davies}(1975)}]{Davies1975}%
		\BibitemOpen
		\bibfield  {author} {\bibinfo {author} {\bibfnamefont {P.~C.~W.}\ \bibnamefont {Davies}},\ }\bibfield  {title} {\bibinfo {title} {Scalar production in {S}chwarzschild and {R}indlermetrics},\ }\href {https://doi.org/10.1088/0305-4470/8/4/022} {\bibfield  {journal} {\bibinfo  {journal} {Journal of Physics A: Mathematical and General}\ }\textbf {\bibinfo {volume} {8}},\ \bibinfo {pages} {609} (\bibinfo {year} {1975})}\BibitemShut {NoStop}%
		\bibitem [{\citenamefont {{U}nruh}(1976)}]{Unruh1976}%
		\BibitemOpen
		\bibfield  {author} {\bibinfo {author} {\bibfnamefont {W.~G.}\ \bibnamefont {{U}nruh}},\ }\bibfield  {title} {\bibinfo {title} {Notes on black-hole evaporation},\ }\href {https://doi.org/10.1103/PhysRevD.14.870} {\bibfield  {journal} {\bibinfo  {journal} {Phys. Rev. D}\ }\textbf {\bibinfo {volume} {14}},\ \bibinfo {pages} {870} (\bibinfo {year} {1976})}\BibitemShut {NoStop}%
		\bibitem [{\citenamefont {Rogers}(1988)}]{Rogers1988}%
		\BibitemOpen
		\bibfield  {author} {\bibinfo {author} {\bibfnamefont {J.}~\bibnamefont {Rogers}},\ }\bibfield  {title} {\bibinfo {title} {Detector for the temperaturelike effect of acceleration},\ }\href {https://doi.org/10.1103/PhysRevLett.61.2113} {\bibfield  {journal} {\bibinfo  {journal} {Phys. Rev. Lett.}\ }\textbf {\bibinfo {volume} {61}},\ \bibinfo {pages} {2113} (\bibinfo {year} {1988})}\BibitemShut {NoStop}%
		\bibitem [{\citenamefont {Chen}\ and\ \citenamefont {Tajima}(1999)}]{Tajima1999}%
		\BibitemOpen
		\bibfield  {author} {\bibinfo {author} {\bibfnamefont {P.}~\bibnamefont {Chen}}\ and\ \bibinfo {author} {\bibfnamefont {T.}~\bibnamefont {Tajima}},\ }\bibfield  {title} {\bibinfo {title} {Testing {U}nruh radiation with ultraintense lasers},\ }\href {https://doi.org/10.1103/PhysRevLett.83.256} {\bibfield  {journal} {\bibinfo  {journal} {Phys. Rev. Lett.}\ }\textbf {\bibinfo {volume} {83}},\ \bibinfo {pages} {256} (\bibinfo {year} {1999})}\BibitemShut {NoStop}%
		\bibitem [{\citenamefont {Vanzella}\ and\ \citenamefont {Matsas}(2001)}]{Vanzella2001}%
		\BibitemOpen
		\bibfield  {author} {\bibinfo {author} {\bibfnamefont {D.~A.~T.}\ \bibnamefont {Vanzella}}\ and\ \bibinfo {author} {\bibfnamefont {G.~E.~A.}\ \bibnamefont {Matsas}},\ }\bibfield  {title} {\bibinfo {title} {Decay of accelerated protons and the existence of the fulling-davies-{U}nruh effect},\ }\href {https://doi.org/10.1103/PhysRevLett.87.151301} {\bibfield  {journal} {\bibinfo  {journal} {Phys. Rev. Lett.}\ }\textbf {\bibinfo {volume} {87}},\ \bibinfo {pages} {151301} (\bibinfo {year} {2001})}\BibitemShut {NoStop}%
		\bibitem [{\citenamefont {Scully}\ \emph {et~al.}(2003)\citenamefont {Scully}, \citenamefont {Kocharovsky}, \citenamefont {Belyanin}, \citenamefont {Fry},\ and\ \citenamefont {Capasso}}]{scully2003}%
		\BibitemOpen
		\bibfield  {author} {\bibinfo {author} {\bibfnamefont {M.~O.}\ \bibnamefont {Scully}}, \bibinfo {author} {\bibfnamefont {V.~V.}\ \bibnamefont {Kocharovsky}}, \bibinfo {author} {\bibfnamefont {A.}~\bibnamefont {Belyanin}}, \bibinfo {author} {\bibfnamefont {E.}~\bibnamefont {Fry}},\ and\ \bibinfo {author} {\bibfnamefont {F.}~\bibnamefont {Capasso}},\ }\bibfield  {title} {\bibinfo {title} {Enhancing acceleration radiation from ground-state atoms via cavity quantum electrodynamics},\ }\href {https://doi.org/10.1103/PhysRevLett.91.243004} {\bibfield  {journal} {\bibinfo  {journal} {Phys. Rev. Lett.}\ }\textbf {\bibinfo {volume} {91}},\ \bibinfo {pages} {243004} (\bibinfo {year} {2003})}\BibitemShut {NoStop}%
		\bibitem [{\citenamefont {Aspachs}\ \emph {et~al.}(2010)\citenamefont {Aspachs}, \citenamefont {Adesso},\ and\ \citenamefont {Fuentes}}]{fuentes2010}%
		\BibitemOpen
		\bibfield  {author} {\bibinfo {author} {\bibfnamefont {M.}~\bibnamefont {Aspachs}}, \bibinfo {author} {\bibfnamefont {G.}~\bibnamefont {Adesso}},\ and\ \bibinfo {author} {\bibfnamefont {I.}~\bibnamefont {Fuentes}},\ }\bibfield  {title} {\bibinfo {title} {Optimal quantum estimation of the {U}nruh-{H}awking effect},\ }\href {https://doi.org/10.1103/PhysRevLett.105.151301} {\bibfield  {journal} {\bibinfo  {journal} {Phys. Rev. Lett.}\ }\textbf {\bibinfo {volume} {105}},\ \bibinfo {pages} {151301} (\bibinfo {year} {2010})}\BibitemShut {NoStop}%
		\bibitem [{\citenamefont {Mart\'{\i}n-Mart\'{\i}nez}\ \emph {et~al.}(2013)\citenamefont {Mart\'{\i}n-Mart\'{\i}nez}, \citenamefont {Montero},\ and\ \citenamefont {del Rey}}]{Martinez2013}%
		\BibitemOpen
		\bibfield  {author} {\bibinfo {author} {\bibfnamefont {E.}~\bibnamefont {Mart\'{\i}n-Mart\'{\i}nez}}, \bibinfo {author} {\bibfnamefont {M.}~\bibnamefont {Montero}},\ and\ \bibinfo {author} {\bibfnamefont {M.}~\bibnamefont {del Rey}},\ }\bibfield  {title} {\bibinfo {title} {Wavepacket detection with the {U}nruh-{D}e{W}itt model},\ }\href {https://doi.org/10.1103/PhysRevD.87.064038} {\bibfield  {journal} {\bibinfo  {journal} {Phys. Rev. D}\ }\textbf {\bibinfo {volume} {87}},\ \bibinfo {pages} {064038} (\bibinfo {year} {2013})}\BibitemShut {NoStop}%
		\bibitem [{\citenamefont {Barshay}\ and\ \citenamefont {Troost}(1978)}]{Barshay1978}%
		\BibitemOpen
		\bibfield  {author} {\bibinfo {author} {\bibfnamefont {S.}~\bibnamefont {Barshay}}\ and\ \bibinfo {author} {\bibfnamefont {W.}~\bibnamefont {Troost}},\ }\bibfield  {title} {\bibinfo {title} {A possible origin for temperature in strong interactions},\ }\href {https://doi.org/https://doi.org/10.1016/0370-2693(78)90759-1} {\bibfield  {journal} {\bibinfo  {journal} {Physics Letters B}\ }\textbf {\bibinfo {volume} {73}},\ \bibinfo {pages} {437} (\bibinfo {year} {1978})}\BibitemShut {NoStop}%
		\bibitem [{\citenamefont {Barshay}\ \emph {et~al.}(1980)\citenamefont {Barshay}, \citenamefont {Braun}, \citenamefont {Gerber},\ and\ \citenamefont {Maurer}}]{Barshay1980}%
		\BibitemOpen
		\bibfield  {author} {\bibinfo {author} {\bibfnamefont {S.}~\bibnamefont {Barshay}}, \bibinfo {author} {\bibfnamefont {H.}~\bibnamefont {Braun}}, \bibinfo {author} {\bibfnamefont {J.~P.}\ \bibnamefont {Gerber}},\ and\ \bibinfo {author} {\bibfnamefont {G.}~\bibnamefont {Maurer}},\ }\bibfield  {title} {\bibinfo {title} {Possible evidence for fluctuations in the hadronic temperature},\ }\href {https://doi.org/10.1103/PhysRevD.21.1849} {\bibfield  {journal} {\bibinfo  {journal} {Phys. Rev. D}\ }\textbf {\bibinfo {volume} {21}},\ \bibinfo {pages} {1849} (\bibinfo {year} {1980})}\BibitemShut {NoStop}%
		\bibitem [{\citenamefont {Kharzeev}(2006)}]{Kharzeev2006}%
		\BibitemOpen
		\bibfield  {author} {\bibinfo {author} {\bibfnamefont {D.}~\bibnamefont {Kharzeev}},\ }\bibfield  {title} {\bibinfo {title} {Quantum black holes and thermalization in relativistic heavy ion collisions},\ }\href {https://doi.org/https://doi.org/10.1016/j.nuclphysa.2006.06.051} {\bibfield  {journal} {\bibinfo  {journal} {Nuclear Physics A}\ }\textbf {\bibinfo {volume} {774}},\ \bibinfo {pages} {315} (\bibinfo {year} {2006})}\BibitemShut {NoStop}%
		\bibitem [{\citenamefont {Salton}\ \emph {et~al.}(2015)\citenamefont {Salton}, \citenamefont {Mann},\ and\ \citenamefont {Menicucci}}]{Salton:2014jaa}%
		\BibitemOpen
		\bibfield  {author} {\bibinfo {author} {\bibfnamefont {G.}~\bibnamefont {Salton}}, \bibinfo {author} {\bibfnamefont {R.~B.}\ \bibnamefont {Mann}},\ and\ \bibinfo {author} {\bibfnamefont {N.~C.}\ \bibnamefont {Menicucci}},\ }\bibfield  {title} {\bibinfo {title} {{Acceleration-assisted entanglement harvesting and rangefinding}},\ }\href {https://doi.org/10.1088/1367-2630/17/3/035001} {\bibfield  {journal} {\bibinfo  {journal} {New J. Phys.}\ }\textbf {\bibinfo {volume} {17}},\ \bibinfo {pages} {035001} (\bibinfo {year} {2015})},\ \Eprint {https://arxiv.org/abs/1408.1395} {arXiv:1408.1395 [quant-ph]} \BibitemShut {NoStop}%
		\bibitem [{\citenamefont {Mart\'{\i}n-Mart\'{\i}nez}\ \emph {et~al.}(2011)\citenamefont {Mart\'{\i}n-Mart\'{\i}nez}, \citenamefont {Fuentes},\ and\ \citenamefont {Mann}}]{Eduardo2011}%
		\BibitemOpen
		\bibfield  {author} {\bibinfo {author} {\bibfnamefont {E.}~\bibnamefont {Mart\'{\i}n-Mart\'{\i}nez}}, \bibinfo {author} {\bibfnamefont {I.}~\bibnamefont {Fuentes}},\ and\ \bibinfo {author} {\bibfnamefont {R.~B.}\ \bibnamefont {Mann}},\ }\bibfield  {title} {\bibinfo {title} {Using {B}erry's phase to detect the {U}nruh effect at lower accelerations},\ }\href {https://doi.org/10.1103/PhysRevLett.107.131301} {\bibfield  {journal} {\bibinfo  {journal} {Phys. Rev. Lett.}\ }\textbf {\bibinfo {volume} {107}},\ \bibinfo {pages} {131301} (\bibinfo {year} {2011})}\BibitemShut {NoStop}%
		\bibitem [{\citenamefont {Hu}\ and\ \citenamefont {Yu}(2012)}]{HuYu2012}%
		\BibitemOpen
		\bibfield  {author} {\bibinfo {author} {\bibfnamefont {J.}~\bibnamefont {Hu}}\ and\ \bibinfo {author} {\bibfnamefont {H.}~\bibnamefont {Yu}},\ }\bibfield  {title} {\bibinfo {title} {Geometric phase for an accelerated two-level atom and the unruh effect},\ }\href {https://doi.org/10.1103/PhysRevA.85.032105} {\bibfield  {journal} {\bibinfo  {journal} {Phys. Rev. A}\ }\textbf {\bibinfo {volume} {85}},\ \bibinfo {pages} {032105} (\bibinfo {year} {2012})}\BibitemShut {NoStop}%
		\bibitem [{\citenamefont {Arya}\ \emph {et~al.}(2022)\citenamefont {Arya}, \citenamefont {Mittal}, \citenamefont {Lochan},\ and\ \citenamefont {Goyal}}]{Arya2022}%
		\BibitemOpen
		\bibfield  {author} {\bibinfo {author} {\bibfnamefont {N.}~\bibnamefont {Arya}}, \bibinfo {author} {\bibfnamefont {V.}~\bibnamefont {Mittal}}, \bibinfo {author} {\bibfnamefont {K.}~\bibnamefont {Lochan}},\ and\ \bibinfo {author} {\bibfnamefont {S.~K.}\ \bibnamefont {Goyal}},\ }\bibfield  {title} {\bibinfo {title} {Geometric phase assisted observation of noninertial cavity-{QED} effects},\ }\href {https://doi.org/10.1103/PhysRevD.106.045011} {\bibfield  {journal} {\bibinfo  {journal} {Phys. Rev. D}\ }\textbf {\bibinfo {volume} {106}},\ \bibinfo {pages} {045011} (\bibinfo {year} {2022})}\BibitemShut {NoStop}%
		\bibitem [{\citenamefont {Arya}\ and\ \citenamefont {Goyal}(2023)}]{Arya2023}%
		\BibitemOpen
		\bibfield  {author} {\bibinfo {author} {\bibfnamefont {N.}~\bibnamefont {Arya}}\ and\ \bibinfo {author} {\bibfnamefont {S.~K.}\ \bibnamefont {Goyal}},\ }\bibfield  {title} {\bibinfo {title} {Lamb shift as a witness for quantum noninertial effects},\ }\href {https://doi.org/10.1103/PhysRevD.108.085011} {\bibfield  {journal} {\bibinfo  {journal} {Phys. Rev. D}\ }\textbf {\bibinfo {volume} {108}},\ \bibinfo {pages} {085011} (\bibinfo {year} {2023})}\BibitemShut {NoStop}%
		\bibitem [{\citenamefont {Lochan}\ \emph {et~al.}(2020)\citenamefont {Lochan}, \citenamefont {Ulbricht}, \citenamefont {Vinante},\ and\ \citenamefont {Goyal}}]{Lochan:2019}%
		\BibitemOpen
		\bibfield  {author} {\bibinfo {author} {\bibfnamefont {K.}~\bibnamefont {Lochan}}, \bibinfo {author} {\bibfnamefont {H.}~\bibnamefont {Ulbricht}}, \bibinfo {author} {\bibfnamefont {A.}~\bibnamefont {Vinante}},\ and\ \bibinfo {author} {\bibfnamefont {S.~K.}\ \bibnamefont {Goyal}},\ }\bibfield  {title} {\bibinfo {title} {{Detecting Acceleration-Enhanced Vacuum Fluctuations with Atoms Inside a Cavity}},\ }\href {https://doi.org/10.1103/PhysRevLett.125.241301} {\bibfield  {journal} {\bibinfo  {journal} {Phys. Rev. Lett.}\ }\textbf {\bibinfo {volume} {125}},\ \bibinfo {pages} {241301} (\bibinfo {year} {2020})},\ \Eprint {https://arxiv.org/abs/1909.09396} {arXiv:1909.09396 [gr-qc]} \BibitemShut {NoStop}%
		\bibitem [{\citenamefont {Vriend}\ \emph {et~al.}(2021)\citenamefont {Vriend}, \citenamefont {Grimmer},\ and\ \citenamefont {Martín-Martínez}}]{Vriend2021}%
		\BibitemOpen
		\bibfield  {author} {\bibinfo {author} {\bibfnamefont {S.}~\bibnamefont {Vriend}}, \bibinfo {author} {\bibfnamefont {D.}~\bibnamefont {Grimmer}},\ and\ \bibinfo {author} {\bibfnamefont {E.}~\bibnamefont {Martín-Martínez}},\ }\bibfield  {title} {\bibinfo {title} {The {U}nruh effect in slow motion},\ }\bibfield  {journal} {\bibinfo  {journal} {Symmetry}\ }\textbf {\bibinfo {volume} {13}},\ \href {https://doi.org/10.3390/sym13111977} {10.3390/sym13111977} (\bibinfo {year} {2021})\BibitemShut {NoStop}%
		\bibitem [{\citenamefont {Stargen}\ and\ \citenamefont {Lochan}(2022)}]{Jaffino2022}%
		\BibitemOpen
		\bibfield  {author} {\bibinfo {author} {\bibfnamefont {D.~J.}\ \bibnamefont {Stargen}}\ and\ \bibinfo {author} {\bibfnamefont {K.}~\bibnamefont {Lochan}},\ }\bibfield  {title} {\bibinfo {title} {Cavity optimization for {U}nruh effect at small accelerations},\ }\href {https://doi.org/10.1103/PhysRevLett.129.111303} {\bibfield  {journal} {\bibinfo  {journal} {Phys. Rev. Lett.}\ }\textbf {\bibinfo {volume} {129}},\ \bibinfo {pages} {111303} (\bibinfo {year} {2022})}\BibitemShut {NoStop}%
		\bibitem [{\citenamefont {Hagley}\ and\ \citenamefont {Pipkin}(1994)}]{Hagley1994}%
		\BibitemOpen
		\bibfield  {author} {\bibinfo {author} {\bibfnamefont {E.~W.}\ \bibnamefont {Hagley}}\ and\ \bibinfo {author} {\bibfnamefont {F.~M.}\ \bibnamefont {Pipkin}},\ }\bibfield  {title} {\bibinfo {title} {Separated oscillatory field measurement of hydrogen 2${\mathit{s}}_{1/2}$-2${\mathit{p}}_{3/2}$ fine structure interval},\ }\href {https://doi.org/10.1103/PhysRevLett.72.1172} {\bibfield  {journal} {\bibinfo  {journal} {Phys. Rev. Lett.}\ }\textbf {\bibinfo {volume} {72}},\ \bibinfo {pages} {1172} (\bibinfo {year} {1994})}\BibitemShut {NoStop}%
		\bibitem [{\citenamefont {Weitz}\ \emph {et~al.}(1994)\citenamefont {Weitz}, \citenamefont {Huber}, \citenamefont {Schmidt-Kaler}, \citenamefont {Leibfried},\ and\ \citenamefont {H\"ansch}}]{Weitz1994}%
		\BibitemOpen
		\bibfield  {author} {\bibinfo {author} {\bibfnamefont {M.}~\bibnamefont {Weitz}}, \bibinfo {author} {\bibfnamefont {A.}~\bibnamefont {Huber}}, \bibinfo {author} {\bibfnamefont {F.}~\bibnamefont {Schmidt-Kaler}}, \bibinfo {author} {\bibfnamefont {D.}~\bibnamefont {Leibfried}},\ and\ \bibinfo {author} {\bibfnamefont {T.~W.}\ \bibnamefont {H\"ansch}},\ }\bibfield  {title} {\bibinfo {title} {Precision measurement of the hydrogen and deuterium 1 s ground state {L}amb shift},\ }\href {https://doi.org/10.1103/PhysRevLett.72.328} {\bibfield  {journal} {\bibinfo  {journal} {Phys. Rev. Lett.}\ }\textbf {\bibinfo {volume} {72}},\ \bibinfo {pages} {328} (\bibinfo {year} {1994})}\BibitemShut {NoStop}%
		\bibitem [{\citenamefont {Berkeland}\ \emph {et~al.}(1995)\citenamefont {Berkeland}, \citenamefont {Hinds},\ and\ \citenamefont {Boshier}}]{Berkeland1995}%
		\BibitemOpen
		\bibfield  {author} {\bibinfo {author} {\bibfnamefont {D.~J.}\ \bibnamefont {Berkeland}}, \bibinfo {author} {\bibfnamefont {E.~A.}\ \bibnamefont {Hinds}},\ and\ \bibinfo {author} {\bibfnamefont {M.~G.}\ \bibnamefont {Boshier}},\ }\bibfield  {title} {\bibinfo {title} {Precise optical measurement of {L}amb shifts in atomic hydrogen},\ }\href {https://doi.org/10.1103/PhysRevLett.75.2470} {\bibfield  {journal} {\bibinfo  {journal} {Phys. Rev. Lett.}\ }\textbf {\bibinfo {volume} {75}},\ \bibinfo {pages} {2470} (\bibinfo {year} {1995})}\BibitemShut {NoStop}%
		\bibitem [{\citenamefont {Bezginov}\ \emph {et~al.}(2019)\citenamefont {Bezginov}, \citenamefont {Valdez}, \citenamefont {Horbatsch}, \citenamefont {Marsman}, \citenamefont {Vutha},\ and\ \citenamefont {Hessels}}]{Bezginov2019}%
		\BibitemOpen
		\bibfield  {author} {\bibinfo {author} {\bibfnamefont {N.}~\bibnamefont {Bezginov}}, \bibinfo {author} {\bibfnamefont {T.}~\bibnamefont {Valdez}}, \bibinfo {author} {\bibfnamefont {M.}~\bibnamefont {Horbatsch}}, \bibinfo {author} {\bibfnamefont {A.}~\bibnamefont {Marsman}}, \bibinfo {author} {\bibfnamefont {A.~C.}\ \bibnamefont {Vutha}},\ and\ \bibinfo {author} {\bibfnamefont {E.~A.}\ \bibnamefont {Hessels}},\ }\bibfield  {title} {\bibinfo {title} {A measurement of the atomic hydrogen {L}amb shift and the proton charge radius},\ }\href {https://doi.org/10.1126/science.aau7807} {\bibfield  {journal} {\bibinfo  {journal} {Science}\ }\textbf {\bibinfo {volume} {365}},\ \bibinfo {pages} {1007} (\bibinfo {year} {2019})}\BibitemShut {NoStop}%
		\bibitem [{\citenamefont {Tiesinga}\ \emph {et~al.}(2021)\citenamefont {Tiesinga}, \citenamefont {Mohr}, \citenamefont {Newell},\ and\ \citenamefont {Taylor}}]{CODATA2018}%
		\BibitemOpen
		\bibfield  {author} {\bibinfo {author} {\bibfnamefont {E.}~\bibnamefont {Tiesinga}}, \bibinfo {author} {\bibfnamefont {P.~J.}\ \bibnamefont {Mohr}}, \bibinfo {author} {\bibfnamefont {D.~B.}\ \bibnamefont {Newell}},\ and\ \bibinfo {author} {\bibfnamefont {B.~N.}\ \bibnamefont {Taylor}},\ }\bibfield  {title} {\bibinfo {title} {{CODATA} recommended values of the fundamental physical constants: 2018},\ }\href {https://doi.org/10.1103/RevModPhys.93.025010} {\bibfield  {journal} {\bibinfo  {journal} {Rev. Mod. Phys.}\ }\textbf {\bibinfo {volume} {93}},\ \bibinfo {pages} {025010} (\bibinfo {year} {2021})}\BibitemShut {NoStop}%
		\bibitem [{\citenamefont {Beyer}\ \emph {et~al.}(2016)\citenamefont {Beyer}, \citenamefont {Maisenbacher}, \citenamefont {Matveev}, \citenamefont {Pohl}, \citenamefont {Khabarova}, \citenamefont {Chang}, \citenamefont {Grinin}, \citenamefont {Lamour}, \citenamefont {Shi}, \citenamefont {Yost}, \citenamefont {Udem}, \citenamefont {H\"{a}nsch},\ and\ \citenamefont {Kolachevsky}}]{Beyer2016}%
		\BibitemOpen
		\bibfield  {author} {\bibinfo {author} {\bibfnamefont {A.}~\bibnamefont {Beyer}}, \bibinfo {author} {\bibfnamefont {L.}~\bibnamefont {Maisenbacher}}, \bibinfo {author} {\bibfnamefont {A.}~\bibnamefont {Matveev}}, \bibinfo {author} {\bibfnamefont {R.}~\bibnamefont {Pohl}}, \bibinfo {author} {\bibfnamefont {K.}~\bibnamefont {Khabarova}}, \bibinfo {author} {\bibfnamefont {Y.}~\bibnamefont {Chang}}, \bibinfo {author} {\bibfnamefont {A.}~\bibnamefont {Grinin}}, \bibinfo {author} {\bibfnamefont {T.}~\bibnamefont {Lamour}}, \bibinfo {author} {\bibfnamefont {T.}~\bibnamefont {Shi}}, \bibinfo {author} {\bibfnamefont {D.~C.}\ \bibnamefont {Yost}}, \bibinfo {author} {\bibfnamefont {T.}~\bibnamefont {Udem}}, \bibinfo {author} {\bibfnamefont {T.~W.}\ \bibnamefont {H\"{a}nsch}},\ and\ \bibinfo {author} {\bibfnamefont {N.}~\bibnamefont {Kolachevsky}},\ }\bibfield  {title} {\bibinfo {title} {Active fiber-based retroreflector providing phase-retracing anti-parallel laser beams for precision spectroscopy},\ }\href
		{https://doi.org/10.1364/OE.24.017470} {\bibfield  {journal} {\bibinfo  {journal} {Opt. Express}\ }\textbf {\bibinfo {volume} {24}},\ \bibinfo {pages} {17470} (\bibinfo {year} {2016})}\BibitemShut {NoStop}%
		\bibitem [{\citenamefont {Beyer}\ \emph {et~al.}(2017)\citenamefont {Beyer}, \citenamefont {Maisenbacher}, \citenamefont {Matveev}, \citenamefont {Pohl}, \citenamefont {Khabarova}, \citenamefont {Grinin}, \citenamefont {Lamour}, \citenamefont {Yost}, \citenamefont {Hänsch}, \citenamefont {Kolachevsky},\ and\ \citenamefont {Udem}}]{Beyer2017}%
		\BibitemOpen
		\bibfield  {author} {\bibinfo {author} {\bibfnamefont {A.}~\bibnamefont {Beyer}}, \bibinfo {author} {\bibfnamefont {L.}~\bibnamefont {Maisenbacher}}, \bibinfo {author} {\bibfnamefont {A.}~\bibnamefont {Matveev}}, \bibinfo {author} {\bibfnamefont {R.}~\bibnamefont {Pohl}}, \bibinfo {author} {\bibfnamefont {K.}~\bibnamefont {Khabarova}}, \bibinfo {author} {\bibfnamefont {A.}~\bibnamefont {Grinin}}, \bibinfo {author} {\bibfnamefont {T.}~\bibnamefont {Lamour}}, \bibinfo {author} {\bibfnamefont {D.~C.}\ \bibnamefont {Yost}}, \bibinfo {author} {\bibfnamefont {T.~W.}\ \bibnamefont {Hänsch}}, \bibinfo {author} {\bibfnamefont {N.}~\bibnamefont {Kolachevsky}},\ and\ \bibinfo {author} {\bibfnamefont {T.}~\bibnamefont {Udem}},\ }\bibfield  {title} {\bibinfo {title} {The {R}ydberg constant and proton size from atomic hydrogen},\ }\href {https://doi.org/10.1126/science.aah6677} {\bibfield  {journal} {\bibinfo  {journal} {Science}\ }\textbf {\bibinfo {volume} {358}},\ \bibinfo {pages} {79} (\bibinfo {year}
			{2017})}\BibitemShut {NoStop}%
		\bibitem [{\citenamefont {Fleurbaey}\ \emph {et~al.}(2018)\citenamefont {Fleurbaey}, \citenamefont {Galtier}, \citenamefont {Thomas}, \citenamefont {Bonnaud}, \citenamefont {Julien}, \citenamefont {Biraben}, \citenamefont {Nez}, \citenamefont {Abgrall},\ and\ \citenamefont {Gu\'ena}}]{Fleurbaey2018}%
		\BibitemOpen
		\bibfield  {author} {\bibinfo {author} {\bibfnamefont {H.}~\bibnamefont {Fleurbaey}}, \bibinfo {author} {\bibfnamefont {S.}~\bibnamefont {Galtier}}, \bibinfo {author} {\bibfnamefont {S.}~\bibnamefont {Thomas}}, \bibinfo {author} {\bibfnamefont {M.}~\bibnamefont {Bonnaud}}, \bibinfo {author} {\bibfnamefont {L.}~\bibnamefont {Julien}}, \bibinfo {author} {\bibfnamefont {F.~m.~c.}\ \bibnamefont {Biraben}}, \bibinfo {author} {\bibfnamefont {F.~m.~c.}\ \bibnamefont {Nez}}, \bibinfo {author} {\bibfnamefont {M.}~\bibnamefont {Abgrall}},\ and\ \bibinfo {author} {\bibfnamefont {J.}~\bibnamefont {Gu\'ena}},\ }\bibfield  {title} {\bibinfo {title} {New measurement of the $1s\ensuremath{-}3s$ transition frequency of hydrogen: Contribution to the proton charge radius puzzle},\ }\href {https://doi.org/10.1103/PhysRevLett.120.183001} {\bibfield  {journal} {\bibinfo  {journal} {Phys. Rev. Lett.}\ }\textbf {\bibinfo {volume} {120}},\ \bibinfo {pages} {183001} (\bibinfo {year} {2018})}\BibitemShut {NoStop}%
		\bibitem [{\citenamefont {Grinin}\ \emph {et~al.}(2020)\citenamefont {Grinin}, \citenamefont {Matveev}, \citenamefont {Yost}, \citenamefont {Maisenbacher}, \citenamefont {Wirthl}, \citenamefont {Pohl}, \citenamefont {Hänsch},\ and\ \citenamefont {Udem}}]{Grinin2020}%
		\BibitemOpen
		\bibfield  {author} {\bibinfo {author} {\bibfnamefont {A.}~\bibnamefont {Grinin}}, \bibinfo {author} {\bibfnamefont {A.}~\bibnamefont {Matveev}}, \bibinfo {author} {\bibfnamefont {D.~C.}\ \bibnamefont {Yost}}, \bibinfo {author} {\bibfnamefont {L.}~\bibnamefont {Maisenbacher}}, \bibinfo {author} {\bibfnamefont {V.}~\bibnamefont {Wirthl}}, \bibinfo {author} {\bibfnamefont {R.}~\bibnamefont {Pohl}}, \bibinfo {author} {\bibfnamefont {T.~W.}\ \bibnamefont {Hänsch}},\ and\ \bibinfo {author} {\bibfnamefont {T.}~\bibnamefont {Udem}},\ }\bibfield  {title} {\bibinfo {title} {Two-photon frequency comb spectroscopy of atomic hydrogen},\ }\href {https://doi.org/10.1126/science.abc7776} {\bibfield  {journal} {\bibinfo  {journal} {Science}\ }\textbf {\bibinfo {volume} {370}},\ \bibinfo {pages} {1061} (\bibinfo {year} {2020})}\BibitemShut {NoStop}%
		\bibitem [{\citenamefont {Heydarizadmotlagh}\ \emph {et~al.}(2024)\citenamefont {Heydarizadmotlagh}, \citenamefont {Skinner}, \citenamefont {Kato}, \citenamefont {George},\ and\ \citenamefont {Hessels}}]{Skinner2024}%
		\BibitemOpen
		\bibfield  {author} {\bibinfo {author} {\bibfnamefont {F.}~\bibnamefont {Heydarizadmotlagh}}, \bibinfo {author} {\bibfnamefont {T.~D.~G.}\ \bibnamefont {Skinner}}, \bibinfo {author} {\bibfnamefont {K.}~\bibnamefont {Kato}}, \bibinfo {author} {\bibfnamefont {M.~C.}\ \bibnamefont {George}},\ and\ \bibinfo {author} {\bibfnamefont {E.~A.}\ \bibnamefont {Hessels}},\ }\bibfield  {title} {\bibinfo {title} {Precision measurement of the $n=2$ triplet $p$ $j=1$ to $j=0$ fine structure of atomic helium using frequency-offset separated oscillatory fields},\ }\href {https://doi.org/10.1103/PhysRevLett.132.163001} {\bibfield  {journal} {\bibinfo  {journal} {Phys. Rev. Lett.}\ }\textbf {\bibinfo {volume} {132}},\ \bibinfo {pages} {163001} (\bibinfo {year} {2024})}\BibitemShut {NoStop}%
		\bibitem [{\citenamefont {Wilson}\ \emph {et~al.}(2003)\citenamefont {Wilson}, \citenamefont {Bushev}, \citenamefont {Eschner}, \citenamefont {Schmidt-Kaler}, \citenamefont {Becher}, \citenamefont {Blatt},\ and\ \citenamefont {Dorner}}]{Wilson2003}%
		\BibitemOpen
		\bibfield  {author} {\bibinfo {author} {\bibfnamefont {M.~A.}\ \bibnamefont {Wilson}}, \bibinfo {author} {\bibfnamefont {P.}~\bibnamefont {Bushev}}, \bibinfo {author} {\bibfnamefont {J.}~\bibnamefont {Eschner}}, \bibinfo {author} {\bibfnamefont {F.}~\bibnamefont {Schmidt-Kaler}}, \bibinfo {author} {\bibfnamefont {C.}~\bibnamefont {Becher}}, \bibinfo {author} {\bibfnamefont {R.}~\bibnamefont {Blatt}},\ and\ \bibinfo {author} {\bibfnamefont {U.}~\bibnamefont {Dorner}},\ }\bibfield  {title} {\bibinfo {title} {Vacuum-field level shifts in a single trapped ion mediated by a single distant mirror},\ }\href {https://doi.org/10.1103/PhysRevLett.91.213602} {\bibfield  {journal} {\bibinfo  {journal} {Phys. Rev. Lett.}\ }\textbf {\bibinfo {volume} {91}},\ \bibinfo {pages} {213602} (\bibinfo {year} {2003})}\BibitemShut {NoStop}%
		\bibitem [{\citenamefont {Marrocco}\ \emph {et~al.}(1998)\citenamefont {Marrocco}, \citenamefont {Weidinger}, \citenamefont {Sang},\ and\ \citenamefont {Walther}}]{Marrocco1998}%
		\BibitemOpen
		\bibfield  {author} {\bibinfo {author} {\bibfnamefont {M.}~\bibnamefont {Marrocco}}, \bibinfo {author} {\bibfnamefont {M.}~\bibnamefont {Weidinger}}, \bibinfo {author} {\bibfnamefont {R.~T.}\ \bibnamefont {Sang}},\ and\ \bibinfo {author} {\bibfnamefont {H.}~\bibnamefont {Walther}},\ }\bibfield  {title} {\bibinfo {title} {Quantum electrodynamic shifts of {R}ydberg energy levels between parallel metal plates},\ }\href {https://doi.org/10.1103/PhysRevLett.81.5784} {\bibfield  {journal} {\bibinfo  {journal} {Phys. Rev. Lett.}\ }\textbf {\bibinfo {volume} {81}},\ \bibinfo {pages} {5784} (\bibinfo {year} {1998})}\BibitemShut {NoStop}%
		\bibitem [{\citenamefont {Heinzen}\ and\ \citenamefont {Feld}(1987)}]{Heinzen1987}%
		\BibitemOpen
		\bibfield  {author} {\bibinfo {author} {\bibfnamefont {D.~J.}\ \bibnamefont {Heinzen}}\ and\ \bibinfo {author} {\bibfnamefont {M.~S.}\ \bibnamefont {Feld}},\ }\bibfield  {title} {\bibinfo {title} {Vacuum radiative level shift and spontaneous-emission linewidth of an atom in an optical resonator},\ }\href {https://doi.org/10.1103/PhysRevLett.59.2623} {\bibfield  {journal} {\bibinfo  {journal} {Phys. Rev. Lett.}\ }\textbf {\bibinfo {volume} {59}},\ \bibinfo {pages} {2623} (\bibinfo {year} {1987})}\BibitemShut {NoStop}%
		\bibitem [{\citenamefont {Karr}\ \emph {et~al.}(2020)\citenamefont {Karr}, \citenamefont {Marchand},\ and\ \citenamefont {Voutier}}]{Karr2020}%
		\BibitemOpen
		\bibfield  {author} {\bibinfo {author} {\bibfnamefont {J.-P.}\ \bibnamefont {Karr}}, \bibinfo {author} {\bibfnamefont {D.}~\bibnamefont {Marchand}},\ and\ \bibinfo {author} {\bibfnamefont {E.}~\bibnamefont {Voutier}},\ }\bibfield  {title} {\bibinfo {title} {The proton size},\ }\href {https://doi.org/10.1038/s42254-020-0229-x} {\bibfield  {journal} {\bibinfo  {journal} {Nature Reviews Physics}\ }\textbf {\bibinfo {volume} {2}},\ \bibinfo {pages} {601} (\bibinfo {year} {2020})}\BibitemShut {NoStop}%
		\bibitem [{\citenamefont {Cohen-Tannoudji}\ \emph {et~al.}(1998)\citenamefont {Cohen-Tannoudji}, \citenamefont {Dupont-Roc},\ and\ \citenamefont {Grynberg}}]{API_Tannoudji}%
		\BibitemOpen
		\bibfield  {author} {\bibinfo {author} {\bibfnamefont {C.}~\bibnamefont {Cohen-Tannoudji}}, \bibinfo {author} {\bibfnamefont {J.}~\bibnamefont {Dupont-Roc}},\ and\ \bibinfo {author} {\bibfnamefont {G.}~\bibnamefont {Grynberg}},\ }\bibinfo {title} {A survey of some interaction processes between photons and atoms},\ in\ \href {https://doi.org/https://doi.org/10.1002/9783527617197.ch2} {\emph {\bibinfo {booktitle} {Atom—Photon Interactions}}}\ (\bibinfo  {publisher} {John Wiley \& Sons, Ltd},\ \bibinfo {year} {1998})\ Chap.~\bibinfo {chapter} {2}, pp.\ \bibinfo {pages} {67--163}\BibitemShut {NoStop}%
		\bibitem [{\citenamefont {Loudon}(2000)}]{loudon2000quantum}%
		\BibitemOpen
		\bibfield  {author} {\bibinfo {author} {\bibfnamefont {R.}~\bibnamefont {Loudon}},\ }\href@noop {} {\emph {\bibinfo {title} {The Quantum Theory of Light}}}\ (\bibinfo  {publisher} {OUP Oxford},\ \bibinfo {year} {2000})\BibitemShut {NoStop}%
		\bibitem [{\citenamefont {Barton}\ and\ \citenamefont {Dalitz}(1970)}]{Barton1970}%
		\BibitemOpen
		\bibfield  {author} {\bibinfo {author} {\bibfnamefont {G.}~\bibnamefont {Barton}}\ and\ \bibinfo {author} {\bibfnamefont {R.~H.}\ \bibnamefont {Dalitz}},\ }\bibfield  {title} {\bibinfo {title} {Quantum electrodynamics of spinless particles between conducting plates},\ }\href {https://doi.org/10.1098/rspa.1970.0208} {\bibfield  {journal} {\bibinfo  {journal} {Proceedings of the Royal Society of London. A. Mathematical and Physical Sciences}\ }\textbf {\bibinfo {volume} {320}},\ \bibinfo {pages} {251} (\bibinfo {year} {1970})}\BibitemShut {NoStop}%
		\bibitem [{\citenamefont {Barton}\ and\ \citenamefont {Dalitz}(1979)}]{Barton1979}%
		\BibitemOpen
		\bibfield  {author} {\bibinfo {author} {\bibfnamefont {G.}~\bibnamefont {Barton}}\ and\ \bibinfo {author} {\bibfnamefont {R.~H.}\ \bibnamefont {Dalitz}},\ }\bibfield  {title} {\bibinfo {title} {The interaction of an atom with electromagnetic vacuum fluctuations in the presence of a pair of perfectly conducting plates},\ }\href {https://doi.org/10.1098/rspa.1979.0079} {\bibfield  {journal} {\bibinfo  {journal} {Proceedings of the Royal Society of London. A. Mathematical and Physical Sciences}\ }\textbf {\bibinfo {volume} {367}},\ \bibinfo {pages} {117} (\bibinfo {year} {1979})}\BibitemShut {NoStop}%
		\bibitem [{\citenamefont {Barton}\ and\ \citenamefont {Dalitz}(1987)}]{Barton1987}%
		\BibitemOpen
		\bibfield  {author} {\bibinfo {author} {\bibfnamefont {G.}~\bibnamefont {Barton}}\ and\ \bibinfo {author} {\bibfnamefont {R.~H.}\ \bibnamefont {Dalitz}},\ }\bibfield  {title} {\bibinfo {title} {Quantum-electrodynamic level shifts between parallel mirrors: analysis},\ }\href {https://doi.org/10.1098/rspa.1987.0032} {\bibfield  {journal} {\bibinfo  {journal} {Proceedings of the Royal Society of London. A. Mathematical and Physical Sciences}\ }\textbf {\bibinfo {volume} {410}},\ \bibinfo {pages} {141} (\bibinfo {year} {1987})}\BibitemShut {NoStop}%
		\bibitem [{\citenamefont {Billaud}\ and\ \citenamefont {Truong}(2012)}]{Billaud2013}%
		\BibitemOpen
		\bibfield  {author} {\bibinfo {author} {\bibfnamefont {B.}~\bibnamefont {Billaud}}\ and\ \bibinfo {author} {\bibfnamefont {T.-T.}\ \bibnamefont {Truong}},\ }\bibfield  {title} {\bibinfo {title} {Lamb shift of non-degenerate energy level systems placed between two infinite parallel conducting plates},\ }\href {https://doi.org/10.1088/1751-8113/46/2/025306} {\bibfield  {journal} {\bibinfo  {journal} {Journal of Physics A: Mathematical and Theoretical}\ }\textbf {\bibinfo {volume} {46}},\ \bibinfo {pages} {025306} (\bibinfo {year} {2012})}\BibitemShut {NoStop}%
		\bibitem [{\citenamefont {{Dalibard, J.}}\ \emph {et~al.}(1984)\citenamefont {{Dalibard, J.}}, \citenamefont {{Dupont-Roc, J.}},\ and\ \citenamefont {{Cohen-Tannoudji, C.}}}]{DDC1984}%
		\BibitemOpen
		\bibfield  {author} {\bibinfo {author} {\bibnamefont {{Dalibard, J.}}}, \bibinfo {author} {\bibnamefont {{Dupont-Roc, J.}}},\ and\ \bibinfo {author} {\bibnamefont {{Cohen-Tannoudji, C.}}},\ }\bibfield  {title} {\bibinfo {title} {Dynamics of a small system coupled to a reservoir : reservoir fluctuations and self-reaction},\ }\href {https://doi.org/10.1051/jphys:01984004504063700} {\bibfield  {journal} {\bibinfo  {journal} {J. Phys. France}\ }\textbf {\bibinfo {volume} {45}},\ \bibinfo {pages} {637} (\bibinfo {year} {1984})}\BibitemShut {NoStop}%
		\bibitem [{\citenamefont {Breuer}\ and\ \citenamefont {Petruccione}(2002)}]{breuer2002}%
		\BibitemOpen
		\bibfield  {author} {\bibinfo {author} {\bibfnamefont {H.-P.}\ \bibnamefont {Breuer}}\ and\ \bibinfo {author} {\bibfnamefont {F.}~\bibnamefont {Petruccione}},\ }\href@noop {} {\emph {\bibinfo {title} {The theory of open quantum systems}}}\ (\bibinfo  {publisher} {Oxford University Press},\ \bibinfo {year} {2002})\BibitemShut {NoStop}%
		\bibitem [{\citenamefont {{D}e{W}itt}(1979)}]{einstein1979general}%
		\BibitemOpen
		\bibfield  {author} {\bibinfo {author} {\bibfnamefont {B.}~\bibnamefont {{D}e{W}itt}},\ }\href@noop {} {\emph {\bibinfo {title} {General relativity: {A}n Einstein {C}entenary {S}urvey}}}\ (\bibinfo  {publisher} {Cambridge University Press},\ \bibinfo {year} {1979})\BibitemShut {NoStop}%
		\bibitem [{\citenamefont {Mart\'{\i}n-Mart\'{\i}nez}\ \emph {et~al.}(2020)\citenamefont {Mart\'{\i}n-Mart\'{\i}nez}, \citenamefont {Perche},\ and\ \citenamefont {de~S.~L.~Torres}}]{Martinez2020}%
		\BibitemOpen
		\bibfield  {author} {\bibinfo {author} {\bibfnamefont {E.}~\bibnamefont {Mart\'{\i}n-Mart\'{\i}nez}}, \bibinfo {author} {\bibfnamefont {T.~R.}\ \bibnamefont {Perche}},\ and\ \bibinfo {author} {\bibfnamefont {B.}~\bibnamefont {de~S.~L.~Torres}},\ }\bibfield  {title} {\bibinfo {title} {General relativistic quantum optics: Finite-size particle detector models in curved spacetimes},\ }\href {https://doi.org/10.1103/PhysRevD.101.045017} {\bibfield  {journal} {\bibinfo  {journal} {Phys. Rev. D}\ }\textbf {\bibinfo {volume} {101}},\ \bibinfo {pages} {045017} (\bibinfo {year} {2020})}\BibitemShut {NoStop}%
		\bibitem [{\citenamefont {Mart\'{\i}n-Mart\'{\i}nez}\ \emph {et~al.}(2021)\citenamefont {Mart\'{\i}n-Mart\'{\i}nez}, \citenamefont {Perche},\ and\ \citenamefont {Torres}}]{Martinez2021}%
		\BibitemOpen
		\bibfield  {author} {\bibinfo {author} {\bibfnamefont {E.}~\bibnamefont {Mart\'{\i}n-Mart\'{\i}nez}}, \bibinfo {author} {\bibfnamefont {T.~R.}\ \bibnamefont {Perche}},\ and\ \bibinfo {author} {\bibfnamefont {B.~d. S.~L.}\ \bibnamefont {Torres}},\ }\bibfield  {title} {\bibinfo {title} {Broken covariance of particle detector models in relativistic quantum information},\ }\href {https://doi.org/10.1103/PhysRevD.103.025007} {\bibfield  {journal} {\bibinfo  {journal} {Phys. Rev. D}\ }\textbf {\bibinfo {volume} {103}},\ \bibinfo {pages} {025007} (\bibinfo {year} {2021})}\BibitemShut {NoStop}%
		\bibitem [{\citenamefont {Audretsch}\ and\ \citenamefont {M\"uller}(1995)}]{Audretsch1995PRA}%
		\BibitemOpen
		\bibfield  {author} {\bibinfo {author} {\bibfnamefont {J.}~\bibnamefont {Audretsch}}\ and\ \bibinfo {author} {\bibfnamefont {R.}~\bibnamefont {M\"uller}},\ }\bibfield  {title} {\bibinfo {title} {Radiative energy shifts of an accelerated two-level system},\ }\href {https://doi.org/10.1103/PhysRevA.52.629} {\bibfield  {journal} {\bibinfo  {journal} {Phys. Rev. A}\ }\textbf {\bibinfo {volume} {52}},\ \bibinfo {pages} {629} (\bibinfo {year} {1995})}\BibitemShut {NoStop}%
		\bibitem [{\citenamefont {Bethe}(1947)}]{Bethe1947}%
		\BibitemOpen
		\bibfield  {author} {\bibinfo {author} {\bibfnamefont {H.~A.}\ \bibnamefont {Bethe}},\ }\bibfield  {title} {\bibinfo {title} {The electromagnetic shift of energy levels},\ }\href {https://doi.org/10.1103/PhysRev.72.339} {\bibfield  {journal} {\bibinfo  {journal} {Phys. Rev.}\ }\textbf {\bibinfo {volume} {72}},\ \bibinfo {pages} {339} (\bibinfo {year} {1947})}\BibitemShut {NoStop}%
		\bibitem [{\citenamefont {Milonni}(1994)}]{Milonni1994}%
		\BibitemOpen
		\bibfield  {author} {\bibinfo {author} {\bibfnamefont {P.~W.}\ \bibnamefont {Milonni}},\ }\href@noop {} {\emph {\bibinfo {title} {The Quantum Vacuum}}}\ (\bibinfo  {publisher} {Academic Press},\ \bibinfo {address} {San Diego},\ \bibinfo {year} {1994})\BibitemShut {NoStop}%
		\bibitem [{\citenamefont {Sakurai}(2006)}]{Sakurai2006}%
		\BibitemOpen
		\bibfield  {author} {\bibinfo {author} {\bibfnamefont {J.~J.}\ \bibnamefont {Sakurai}},\ }\href@noop {} {\emph {\bibinfo {title} {Advanced quantum mechanics}}}\ (\bibinfo  {publisher} {Pearson Education India},\ \bibinfo {year} {2006})\BibitemShut {NoStop}%
		\bibitem [{Note1()}]{Note1}%
		\BibitemOpen
		\bibinfo {note} {The transition rates of an atom undergoing uniform linear acceleration can show thermal character even if the restricted field state is not thermal. See Refs.~\cite {Rovelli2012,Carballo2019}}\BibitemShut {NoStop}%
		\bibitem [{\citenamefont {Rindler}(2006)}]{RindlerRel}%
		\BibitemOpen
		\bibfield  {author} {\bibinfo {author} {\bibfnamefont {W.}~\bibnamefont {Rindler}},\ }\href@noop {} {\emph {\bibinfo {title} {Essential Relativity: Special, General, and Cosmological}}},\ \bibinfo {edition} {2nd}\ ed.\ (\bibinfo  {publisher} {Oxford University Press, NY},\ \bibinfo {year} {2006})\BibitemShut {NoStop}%
		\bibitem [{\citenamefont {Kubo}(1957)}]{Kubo1957}%
		\BibitemOpen
		\bibfield  {author} {\bibinfo {author} {\bibfnamefont {R.}~\bibnamefont {Kubo}},\ }\bibfield  {title} {\bibinfo {title} {Statistical-mechanical theory of irreversible processes. {I}. general theory and simple applications to magnetic and conduction problems},\ }\href {https://doi.org/10.1143/JPSJ.12.570} {\bibfield  {journal} {\bibinfo  {journal} {Journal of the Physical Society of Japan}\ }\textbf {\bibinfo {volume} {12}},\ \bibinfo {pages} {570} (\bibinfo {year} {1957})}\BibitemShut {NoStop}%
		\bibitem [{\citenamefont {Martin}\ and\ \citenamefont {Schwinger}(1959)}]{Martin-Schwinger1959}%
		\BibitemOpen
		\bibfield  {author} {\bibinfo {author} {\bibfnamefont {P.~C.}\ \bibnamefont {Martin}}\ and\ \bibinfo {author} {\bibfnamefont {J.}~\bibnamefont {Schwinger}},\ }\bibfield  {title} {\bibinfo {title} {Theory of many-particle systems. {I}},\ }\href {https://doi.org/10.1103/PhysRev.115.1342} {\bibfield  {journal} {\bibinfo  {journal} {Phys. Rev.}\ }\textbf {\bibinfo {volume} {115}},\ \bibinfo {pages} {1342} (\bibinfo {year} {1959})}\BibitemShut {NoStop}%
		\bibitem [{\citenamefont {Lidar}(2019)}]{Lidar2019notes}%
		\BibitemOpen
		\bibfield  {author} {\bibinfo {author} {\bibfnamefont {D.~A.}\ \bibnamefont {Lidar}},\ }\bibfield  {title} {\bibinfo {title} {Lecture notes on the theory of open quantum systems},\ }\href@noop {} {\bibfield  {journal} {\bibinfo  {journal} {arXiv preprint arXiv:1902.00967}\ } (\bibinfo {year} {2019})}\BibitemShut {NoStop}%
		\bibitem [{\citenamefont {Ju\'arez-Aubry}\ and\ \citenamefont {Moustos}(2019)}]{Aubry2019}%
		\BibitemOpen
		\bibfield  {author} {\bibinfo {author} {\bibfnamefont {B.~A.}\ \bibnamefont {Ju\'arez-Aubry}}\ and\ \bibinfo {author} {\bibfnamefont {D.}~\bibnamefont {Moustos}},\ }\bibfield  {title} {\bibinfo {title} {Asymptotic states for stationary {U}nruh-{D}e{W}itt detectors},\ }\href {https://doi.org/10.1103/PhysRevD.100.025018} {\bibfield  {journal} {\bibinfo  {journal} {Phys. Rev. D}\ }\textbf {\bibinfo {volume} {100}},\ \bibinfo {pages} {025018} (\bibinfo {year} {2019})}\BibitemShut {NoStop}%
		\bibitem [{\citenamefont {Jeffrey}\ \emph {et~al.}(2007)\citenamefont {Jeffrey}, \citenamefont {Zwillinger}, \citenamefont {Gradshteyn},\ and\ \citenamefont {Ryzhik}}]{Gradshteyn}%
		\BibitemOpen
		\bibinfo {editor} {\bibfnamefont {A.}~\bibnamefont {Jeffrey}}, \bibinfo {editor} {\bibfnamefont {D.}~\bibnamefont {Zwillinger}}, \bibinfo {editor} {\bibfnamefont {I.}~\bibnamefont {Gradshteyn}},\ and\ \bibinfo {editor} {\bibfnamefont {I.}~\bibnamefont {Ryzhik}},\ eds.,\ \href {https://doi.org/https://doi.org/10.1016/B978-0-08-047111-2.50016-9} {\emph {\bibinfo {title} {Table of Integrals, Series, and Products (Seventh Edition)}}}\ (\bibinfo  {publisher} {Academic Press},\ \bibinfo {address} {Boston},\ \bibinfo {year} {2007})\BibitemShut {NoStop}%
		\bibitem [{\citenamefont {Heitler}(1954)}]{Heitler1954}%
		\BibitemOpen
		\bibfield  {author} {\bibinfo {author} {\bibfnamefont {W.}~\bibnamefont {Heitler}},\ }\href@noop {} {\emph {\bibinfo {title} {The quantum theory of radiation}}}\ (\bibinfo  {publisher} {Dover Publications},\ \bibinfo {year} {1954})\ p.~\bibinfo {pages} {69}\BibitemShut {NoStop}%
		\bibitem [{\citenamefont {Crispino}\ \emph {et~al.}(2008)\citenamefont {Crispino}, \citenamefont {Higuchi},\ and\ \citenamefont {Matsas}}]{Crispino2008}%
		\BibitemOpen
		\bibfield  {author} {\bibinfo {author} {\bibfnamefont {L.~C.~B.}\ \bibnamefont {Crispino}}, \bibinfo {author} {\bibfnamefont {A.}~\bibnamefont {Higuchi}},\ and\ \bibinfo {author} {\bibfnamefont {G.~E.~A.}\ \bibnamefont {Matsas}},\ }\bibfield  {title} {\bibinfo {title} {The {U}nruh effect and its applications},\ }\href {https://doi.org/10.1103/RevModPhys.80.787} {\bibfield  {journal} {\bibinfo  {journal} {Rev. Mod. Phys.}\ }\textbf {\bibinfo {volume} {80}},\ \bibinfo {pages} {787} (\bibinfo {year} {2008})}\BibitemShut {NoStop}%
		\bibitem [{\citenamefont {Harris}(2008)}]{Harris2008}%
		\BibitemOpen
		\bibfield  {author} {\bibinfo {author} {\bibfnamefont {F.~E.}\ \bibnamefont {Harris}},\ }\bibfield  {title} {\bibinfo {title} {Incomplete bessel, generalized incomplete gamma, or leaky aquifer functions},\ }\href {https://doi.org/https://doi.org/10.1016/j.cam.2007.04.008} {\bibfield  {journal} {\bibinfo  {journal} {Journal of Computational and Applied Mathematics}\ }\textbf {\bibinfo {volume} {215}},\ \bibinfo {pages} {260} (\bibinfo {year} {2008})}\BibitemShut {NoStop}%
		\bibitem [{\citenamefont {Olver}(1997)}]{Olver1997}%
		\BibitemOpen
		\bibfield  {author} {\bibinfo {author} {\bibfnamefont {F.}~\bibnamefont {Olver}},\ }\href@noop {} {\emph {\bibinfo {title} {Asymptotics and special functions}}}\ (\bibinfo  {publisher} {AK Peters/CRC Press},\ \bibinfo {year} {1997})\BibitemShut {NoStop}%
		\bibitem [{\citenamefont {Rovelli}\ and\ \citenamefont {Smerlak}(2012)}]{Rovelli2012}%
		\BibitemOpen
		\bibfield  {author} {\bibinfo {author} {\bibfnamefont {C.}~\bibnamefont {Rovelli}}\ and\ \bibinfo {author} {\bibfnamefont {M.}~\bibnamefont {Smerlak}},\ }\bibfield  {title} {\bibinfo {title} {{U}nruh effect without trans-horizon entanglement},\ }\href {https://doi.org/10.1103/PhysRevD.85.124055} {\bibfield  {journal} {\bibinfo  {journal} {Phys. Rev. D}\ }\textbf {\bibinfo {volume} {85}},\ \bibinfo {pages} {124055} (\bibinfo {year} {2012})}\BibitemShut {NoStop}%
		\bibitem [{\citenamefont {Carballo-Rubio}\ \emph {et~al.}(2019)\citenamefont {Carballo-Rubio}, \citenamefont {Garay}, \citenamefont {Mart\'{\i}n-Mart\'{\i}nez},\ and\ \citenamefont {de~Ram\'on}}]{Carballo2019}%
		\BibitemOpen
		\bibfield  {author} {\bibinfo {author} {\bibfnamefont {R.}~\bibnamefont {Carballo-Rubio}}, \bibinfo {author} {\bibfnamefont {L.~J.}\ \bibnamefont {Garay}}, \bibinfo {author} {\bibfnamefont {E.}~\bibnamefont {Mart\'{\i}n-Mart\'{\i}nez}},\ and\ \bibinfo {author} {\bibfnamefont {J.}~\bibnamefont {de~Ram\'on}},\ }\bibfield  {title} {\bibinfo {title} {{U}nruh effect without thermality},\ }\href {https://doi.org/10.1103/PhysRevLett.123.041601} {\bibfield  {journal} {\bibinfo  {journal} {Phys. Rev. Lett.}\ }\textbf {\bibinfo {volume} {123}},\ \bibinfo {pages} {041601} (\bibinfo {year} {2019})}\BibitemShut {NoStop}%
	\end{thebibliography}
\end{document}